\documentclass{aa}
\usepackage{subfigure}
\usepackage{supertabular}
\usepackage{lscape}
\usepackage{color, soul}

\usepackage{graphicx}
\usepackage{txfonts}
\begin{document}

   \title{Testing cosmic anisotropy with Pantheon sample and quasars at high redshifts}

  % \subtitle{I. Overviewing the $\kappa$-mechanism}

   \author{J. P. Hu\inst{1} 
          \and Y. Y. Wang\inst{1} 
          \and F. Y. Wang\inst{1,2}
          }

   \institute{School of Astronomy and Space Science, Nanjing University, Nanjing 210093, China\\
             \email{fayinwang@nju.edu.cn}
         \and
             Key Laboratory of Modern Astronomy and Astrophysics (Nanjing University), Ministry of Education, Nanjing 210093, China
             }

   \date{Received date; accepted date}

% \abstract{} 
% 5 {} token are mandatory
 
  \abstract
   {In this paper, we investigate the cosmic anisotropy from the SN-Q sample,
        consisting of the Pantheon sample and quasars, by employing
        the hemisphere comparison (HC) method and the dipole fitting (DF)
        method. Compared to the Pantheon sample, the new sample has a larger
        redshift range, a more homogeneous distribution, and a larger sample
        size. For the HC method, we find that the maximum anisotropy level
        is $AL_{max}=0.142\pm0.026$ in the direction
        ($l$, $b$) = $({316.08^{\circ}}^{+27.41}_{-129.48}$, ${4.53^{\circ}}^{+26.29}_{-64.06})$. The magnitude of anisotropy is
        $A$ = ($-$8.46 $^{+4.34}_{-5.51}$)$\times$$10^{-4}$ and the corresponding preferred direction points
        toward $(l$, $b)$ = ($29.31^{\circ}$$^{+30.59}_{-30.54}$, $71.40^{\circ}$$^{+9.79}_{-9.72}$) for the quasar
        sample from the DF method. The combined SN and quasar sample is consistent with the isotropy hypothesis. The distribution of the dataset might impact the preferred direction from the dipole results. The result is weakly dependent on the redshift from the redshift tomography analysis. There
        is no evidence of cosmic anisotropy in the SN-Q sample. Though some results obtained from the quasar sample are not consistent with the standard cosmological model, we still do not find any distinct evidence of cosmic anisotropy in the SN-Q sample.}
           \maketitle
%
%-------------------------------------------------------------------

\section{Introduction}

   The Lambda cold dark matter model ($\rm \Lambda$CDM) is consistent with
    most astronomical observations
   \citep{2010MNRAS.404...60R,2011ApJ...742...16T,2013ApJS..208...20B,2013ApJS..208...19H,2014AA...571A..26P}.
   The basis of the $\rm \Lambda$CDM model assumes that the universe is
   homogeneous and isotropic on a large scale
   \citep{1972Wiley,2008Oxford}. This hypothsis is known as the cosmological principle, and \cite{2010CQG.27.124008C} discuss the requirements in order to probe it. But several analyses of observations
   indicate that the universe may be anisotropic, for instance, these include quasar
   polarization vectors \citep{2005AA...441..915H}, the fine-structure
   constant \citep{2011PhRvL.107s1101W,2012MNRAS.422.3370K}, the
   direct measure of the Hubble parameter \citep{2006PhRvL..96s1302B}, the
   anisotropic dark energy \citep{2008JCAP...06..018K}, the cosmic
   microwave background (CMB)
   \citep{2003PhRvD..68l3523T,2004MNRAS.355.1283B,2004ApJ...605...14E,2010ApJ...714L.265K,2011MNRAS.411.1445G,2014PhRvD..89b3010Z,2015MNRAS.449.3458C,2019Arxiv...1906.02552A}, the large dipole of radio source counts \citep{2011ApJ..742...L23S, 2012MNRAS.427.1994G,2013AA...555..A117R,2016JCAP...03...062T,2017MNRAS.471.1045C,2018JCAP...04...031B,2019PhRvD..100...063501S}, the quasar vector polarization aligment \citep{2016AA...590..A53P,2019AA...622...A113T}, and the galaxy number counts in optical and IR wavelengths \citep{2015MNRAS...449..670A,2017AA...597...A120J,2018MNRAS...475...L106B}. 
   These works hint that the universe may have a preferred expanding
   direction \citep{Perivolaropoulos14}.
   
   In recent years, type Ia supernovae (SNe Ia)
   \citep{2010ApJ...716..712A,2012ApJ...746...85S,2014AA...568A..22B,2018ApJ...859..101S}
   have been widely employed to test cosmic isotropy.
   \cite{2010JCAP...12..012A} searched for the preferred direction of
   anisotropy for the Union2 sample by adopting the hemisphere
   comparison (HC) method \citep{2007AA...474..717S}. They found a
   maximum accelerating expansion rate, which corresponds to a preferred
   direction of anisotropy. After that, Mariano and Perivolaropoulos \citep{2012PhRvD..86h3517M}
   found a possible preferred anisotropic direction at the 2$\sigma$ level
   using the Union2 sample, but by employing the dipole fitting (DF) method.
   Since then, these two methods have been widely used to explore the
   cosmic anisotropy
   \citep{2012JCAP...02..004C,2013PhRvD..87l3522C,2013IJMPD..2250060Z,2013EPJC...73.2653L,2014MNRAS.439.1855H,2015ApJ...808...39B,2018PhRvD..97h3518A,2019EPJC...79..783S}
   by investigating observational data of, for instance, the Union2.1 sample
   \citep{2014MNRAS.437.1840Y,2015ApJ...810...47J,2016MNRAS.460..617L},
the Joint Light-Curve Analysis (JLA) sample
   \citep{2016MNRAS.456.1881L,2018MNRAS.478.3633C,2018MNRAS.474.3516W},
   the Pantheon sample \citep{2018MNRAS.478.5153S}, gamma-ray bursts \citep[GRBs;][]{2014MNRAS.443.1680W}, galaxies
   \citep{2017ApJ...847...86Z}, as well as gravitational wave and fast radio bursts
   \citep{2019arXiv190203580Q,2019JCAP...09..016C}. Using HC and DF
   methods, \cite{2019MNRAS.486.5679Z} studied the cosmic anisotropy
   via the Pantheon sample. They found that the SDSS sample plays a
   decisive role in the Pantheon sample. It may imply that the
   inhomogeneous distribution has a significant effect on the cosmic
   anisotropy \citep{2018ChPhC..42k5103C}. This opinion was also
   presented by \cite{2019EPJC...79..783S}. Their conclusions show
   that the effect of redshift on the result is weak and there is a
   negligible anisotropy when making a redshift tomography.
   \cite{2018EPJC...78..755D} tested the cosmic anisotropy with
   the Pantheon sample, but by using the following three methods: the HC method, the DF method, and 
   Healpix\footnote{http://healpix.sourceforge.net} \citep{2005ApJ...622..759G}.\ They also
    performed a cross check. There are two preferred directions from the HC
   method. In adopting the DF method and Healpix, they found no noticeable
   anisotropy. They also compared the HC method with the DF method by using the
   JLA sample \citep{2018PhRvD..97l3515D} and found that the results of
   these two methods have not always been approximately coincident with
   each other. In order to better test the cosmic isotropy, the best
   way would be to add new samples with a relatively homogeneous distribution.
   
   Quasars can be regarded as quasi-standard candles
   \citep{2017AA...602A..79L,2019AA...631A.120S} by
   using the nonlinear relation
   \citep{1986ApJ...305...83A,2016ApJ...819..154L} between the UV and
   X-ray monochromatic luminosities. Thus, the nonlinear relation can be used for
   cosmological purposes
   \citep{1989SoSAO..61..101K,1997ApSS.253...19Q,2015ApJ...815...33R,2017FrASS...4...68B,2019MNRAS.489..517M,2020MNRAS492.4456K,2020PhRvD101.043502V,2020ApJ888...99W,2020FrASS...7....8L}.
   The quasar sample that consists of 1598 sources was built by
   \cite{2019NatAs...3..272R}. They employed this sample to verify cosmological constraints. \citep{2019AA...628L...4L} and found that a
   deviation from the standard cosmological model emerges, with a
   statistical significance of $~4\sigma$. Whether this deviation will
   appear in cosmic isotropy is unclear. In this paper, we test the cosmological anisotropy from the SN-Q sample,
   which consists of the Pantheon sample and the quasar sample, by adopting
   the HC and DF methods. A fiducial value of $H_{0}$ = 70 km s$^{-1}$ Mpc$^{-1}$ is adopted in this paper. The structure of the paper is as follows. In
   Section 2, we present the quasar sample and discuss the advantages
   of the SN-Q sample over the single Pantheon sample. The best-fitting
   values of the quasar and SN-Q samples are obtained by using the Markov chain Monte Carlo (MCMC) method.
   In Section 3, the HC and DF methods are given. By using these two
   methods, we investigate the cosmic anisotropy by employing the SN-Q
   sample. Finally, a brief summary is presented in the
   last section.

%--------------------------------------------------------------------
\section{The observational data and MCMC fitting}

\subsection{The SNe Ia sample and the quasar sample}
\label{sec:sample}
\begin{figure*}
        \centering
        \includegraphics[width=0.75\hsize]{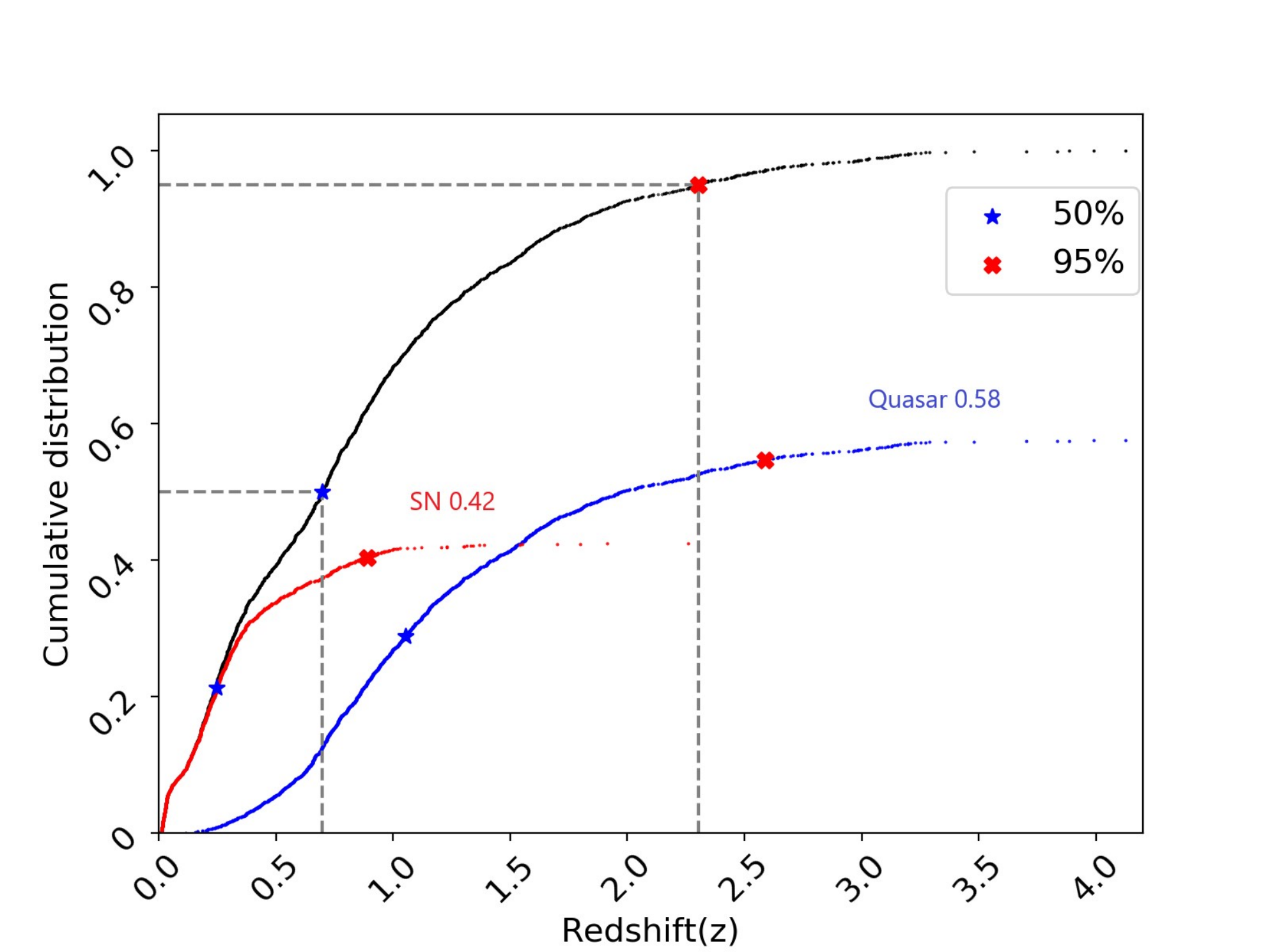}
        \caption{Redshift cumulative distributions. Red, blue, and black lines correspond to the Pantheon, quasar, and SN-Q samples, respectively. The SNe Ia account for 42\% of the SN-Q sample and the quasars for 58\%.}
        \label{F1}       % Give a unique label
\end{figure*}

\begin{figure*}
        \centering
        \subfigure[Pantheon sample]{\label{Fig1.sub.1}\includegraphics[width=0.43\hsize]{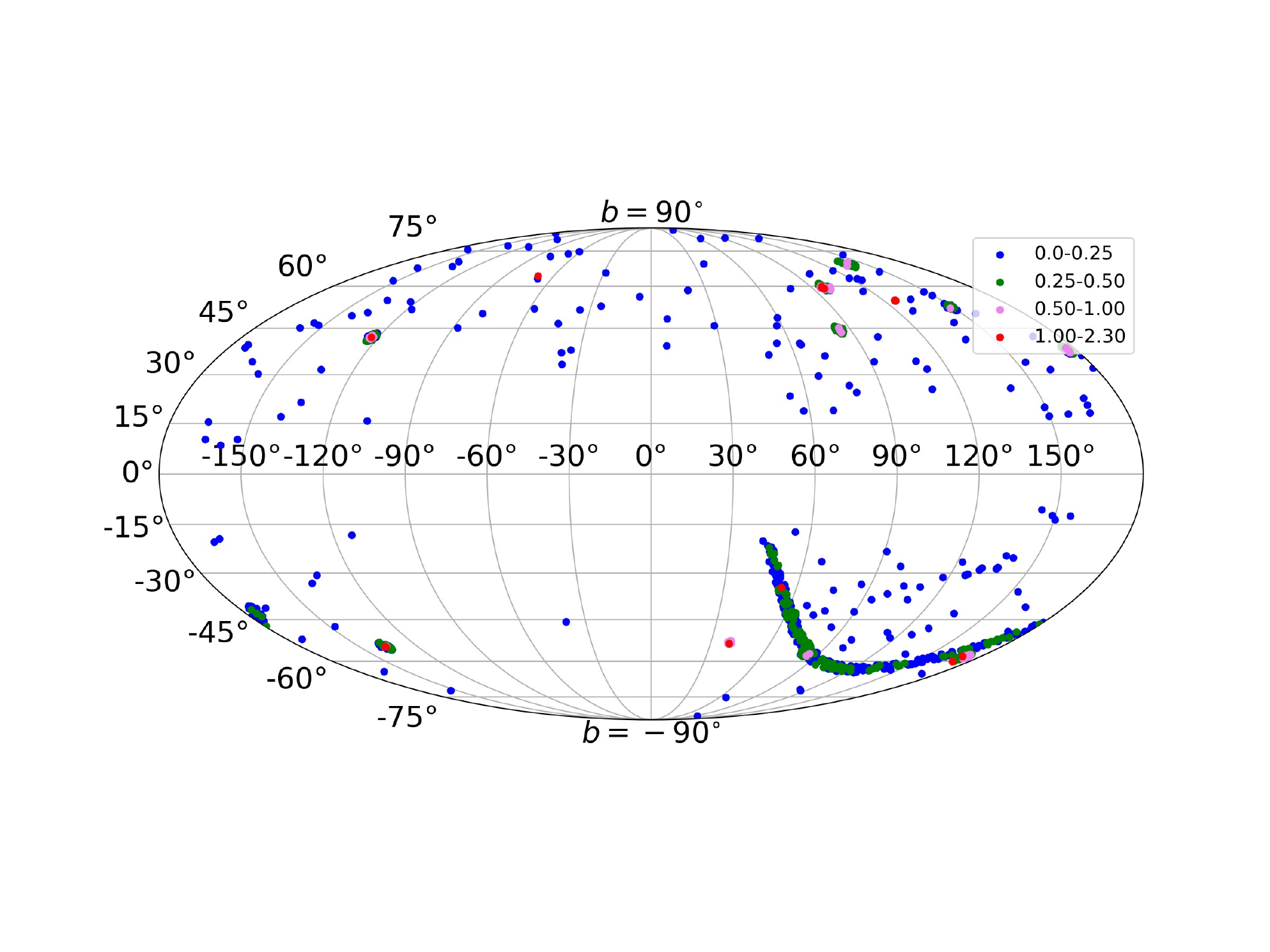}}
        \subfigure[Pantheon sample]{\label{Fig1.sub.2}\includegraphics[width=0.43\hsize]{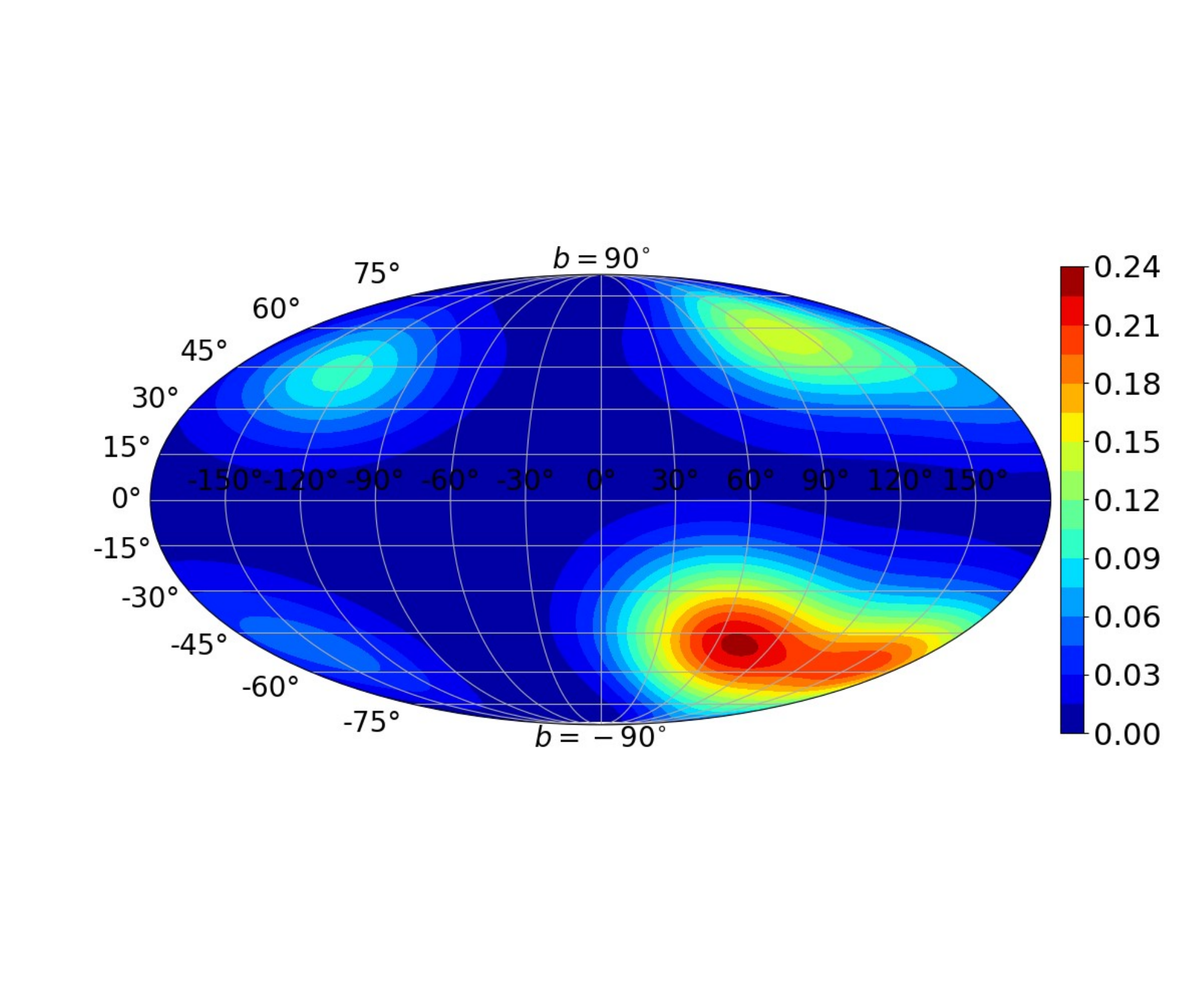}}\\
        \subfigure[The quasar sample]{
                \label{Fig1.sub.3}\includegraphics[width=0.43\hsize]{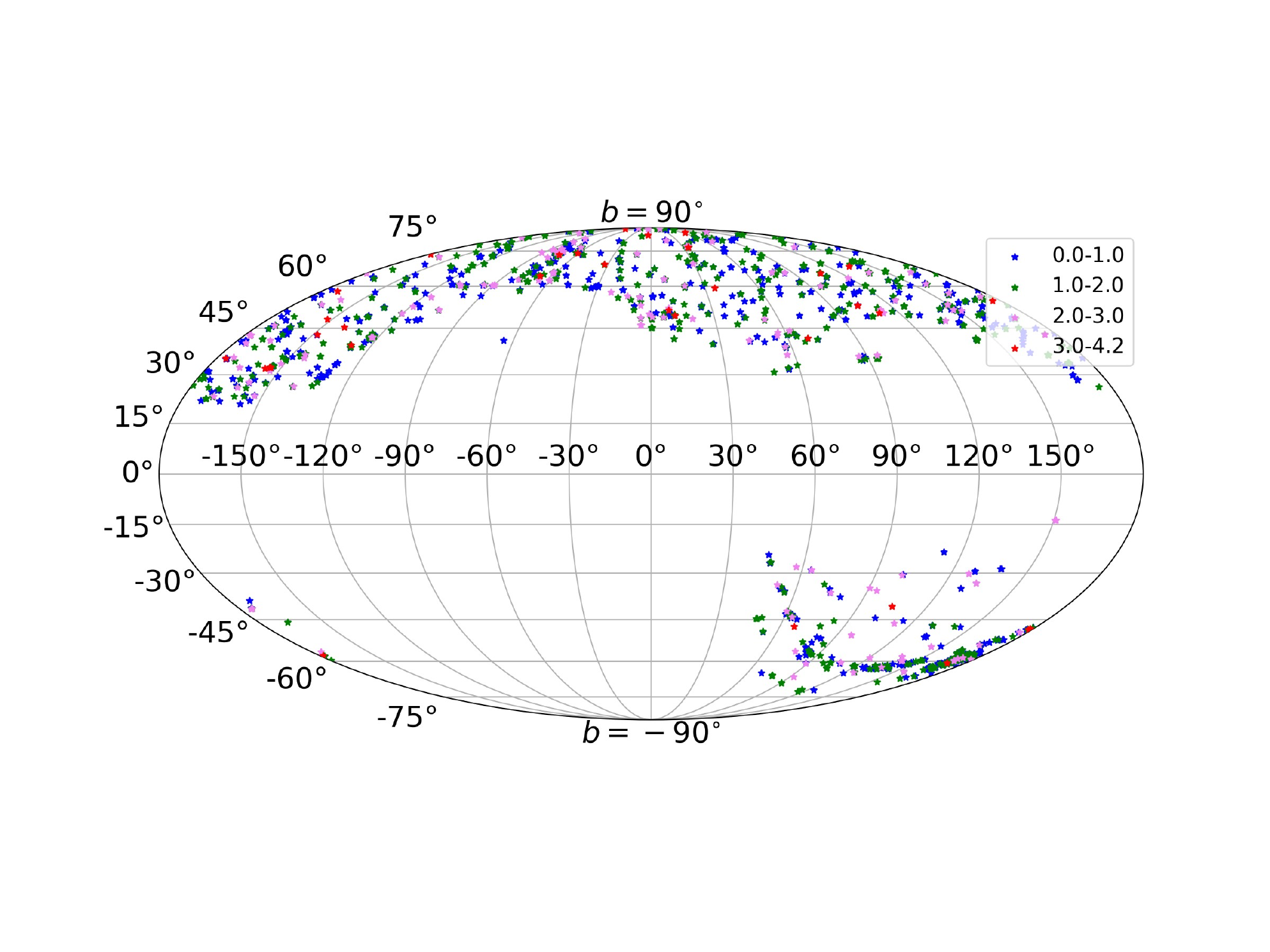}}
        \subfigure[The quasar sample]{
                \label{Fig1.sub.4}\includegraphics[width=0.43\hsize]{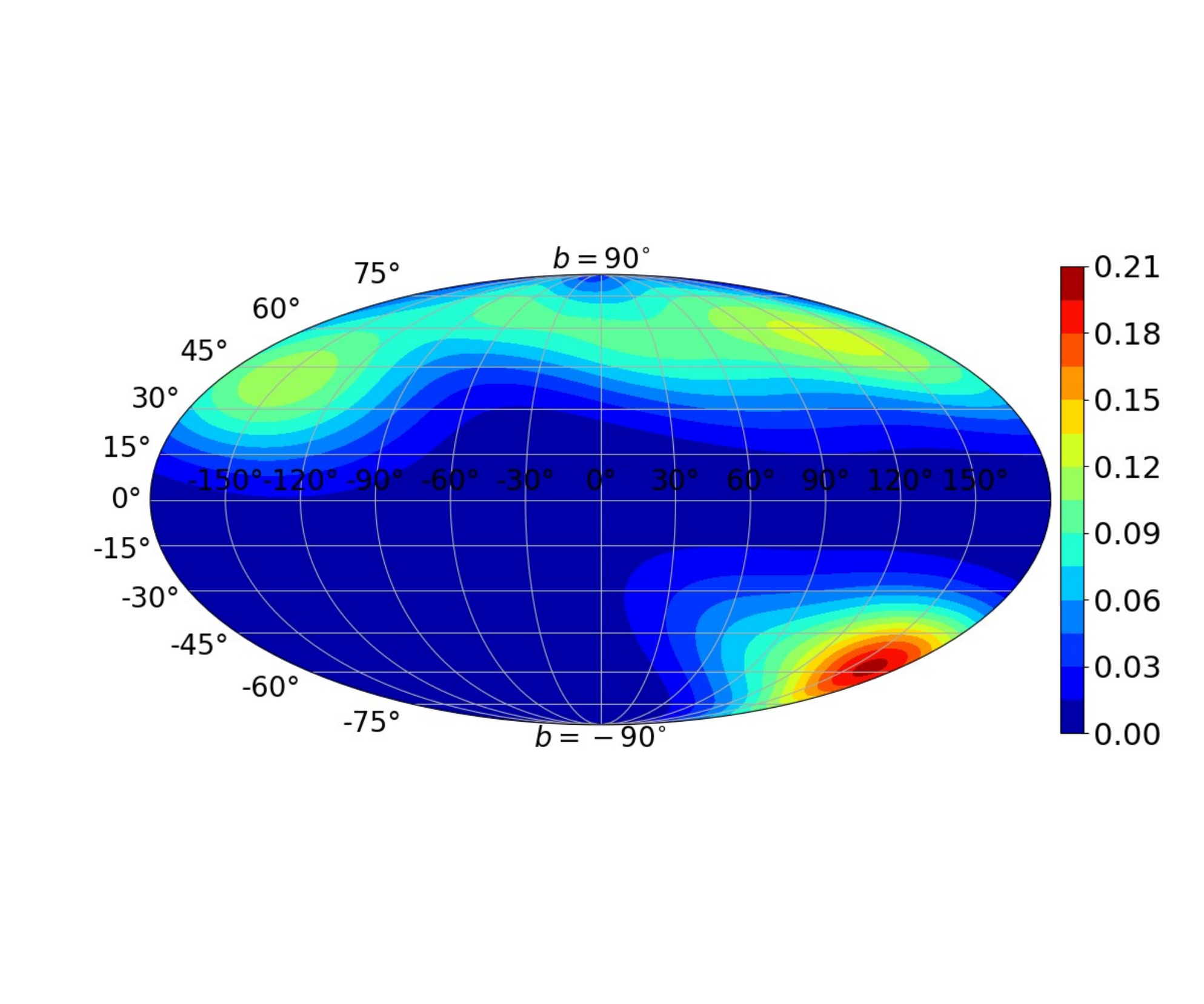}}\\
        \subfigure[SN-Q sample]{
                \label{Fig1.sub.5}\includegraphics[width=0.43\hsize]{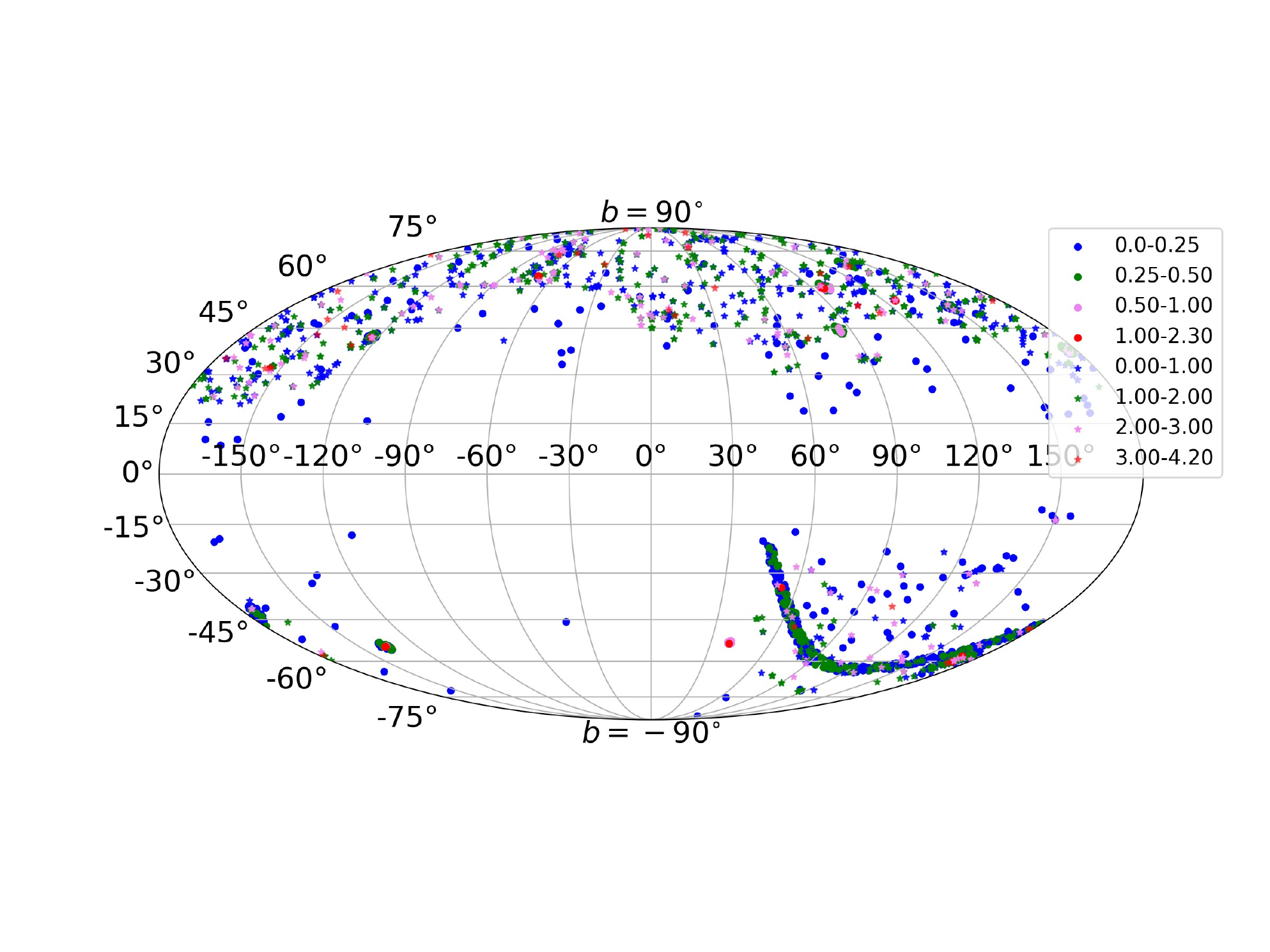}}
        \subfigure[SN-Q sample]{
                \label{Fig1.sub.6}\includegraphics[width=0.43\hsize]{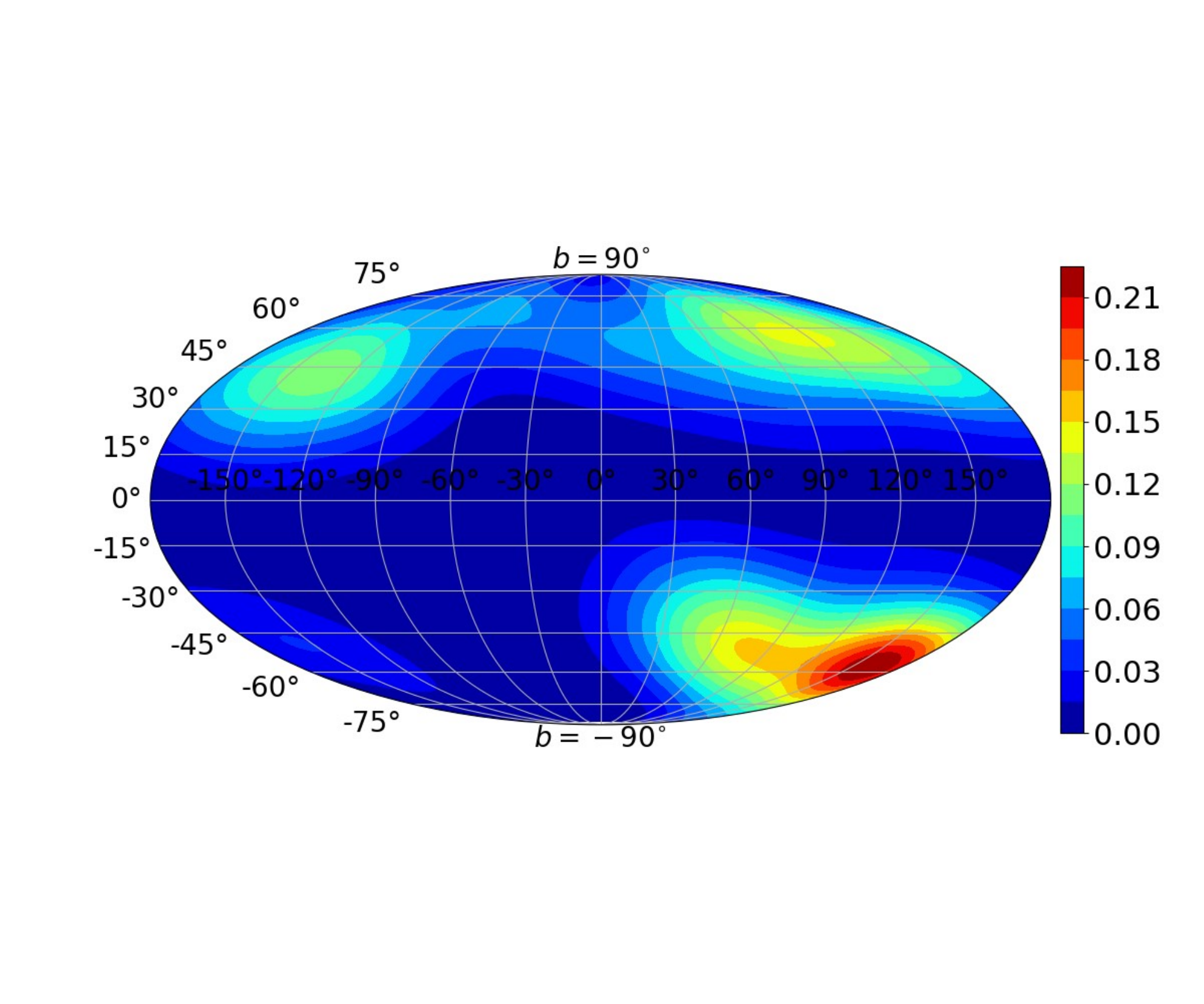}}
        \caption{Distributions and density contours in the galactic coordinate system. Panels (a), (c), and (e) are the coordinate distributions of the Pantheon, quasar, and SN-Q samples in the galactic coordinate system, respectively. The corresponding density contours are described in panels (b), (d), and (f).}
        \label{F2}       % Give a unique label
\end{figure*}

In this paper, we adopt the MCMC method provided by
$emcee$\footnote{https://emcee.readthedocs.io/en/stable/} to
explore the whole parameter space and investigate the cosmic
anisotropy by using the Pantheon and quasar samples. The Pantheon sample,
which was compiled by \cite{2018ApJ...859..101S}, contains 1048 SNe Ia
covering the redshift range of 0.01 $<$ $z$ $<$ 2.30. The quasar
sample consists of 1598 quasar sources \citep{2019NatAs...3..272R}:
1403 sources from SDSS DR7 \citep{2011ApJS..194...45S} and DR12
\citep{2017AA...597A..79P}, 102 from XMM-COSMOS
\citep{2010AA...512A..34L,2010ApJ...716..348B}, 19 from ChaMP
\citep{2014MNRAS.445.1430K}, and 74 from other quasars
\citep{2011ApJS..194...45S,2014MNRAS.445.1430K,2013AA...550A..71V}.
Compared with the Pantheon sample, the quasar sample has a wider range
of redshifts from 0.04 $<$ $z$ $<$ 5.10 and a larger size. Since some
quasars have no coordinate information, we selected 1421 quasars as
the final quasar sample, which covers a redshift of 0.11 $<$ $z$ $<$
4.13. Combining the Pantheon and quasar samples, we obtain a new
sample named the SN-Q sample. In Figure \ref{F1}, the redshift
cumulative distributions of the Pantheon, quasar, and SN-Q samples
are given. In assessing the curve behaviors in Figure \ref{F1}, we find
that supernovae and
quasars occupy forty-two percent and fifty-eight percent in the
total sample, respectively. Ninety-five percent of the SN-Q sample
distributes within $z$ = $2.3$. The distributions and corresponding
densities of these three samples in the galactic coordinates are
illustrated in Figure \ref{F2}. 

The cosmic anisotropy obtained from the SNe sample might rely on
the inhomogeneous distribution of SNe Ia. The distributions of SNe Ia from the Pantheon sample are shown in panel (a) of Figure \ref{F2}. It is obvious that the belt part (the SDSS sample) plays a major role in the full Pantheon sample. To make it easier to comprehend this focus, we plotted the density distribution of the Pantheon sample, which is shown in panel (b) of Figure \ref{F2}. \cite{2019MNRAS.486.5679Z} analyzed the effect of the inhomogeneous distribution of the Pantheon sample on the cosmic anisotropy and found that the SDSS sample plays the most important role in the Pantheon sample. From panels (c) and (d) of
Figure \ref{F2}, we find that the distribution of the quasar
sample is more homogeneous and its maximum density value is smaller
than that of the Pantheon sample. Compared with the Pantheon sample,
adding the quasar sample reduces the unevenness of the overall
sample and weakens the decisive role of the SDSS subsample as shown in panels (e) and (f) of Figure \ref{F2}. By combining Figures \ref{F1} and \ref{F2}, we find that the SN-Q sample is better than the Pantheon sample in terms of quantity, redshift range, and
the uniform of distribution. However, the unevenness may not have been eliminated. It is obvious that the data number of the North Galactic Hemisphere is larger than the South Galactic Hemisphere in the quasar sample. Through statistics, the data number of the South and North Galactic Hemispheres in the Pantheon sample, quasar sample, and SN-Q sample are (644, 404), (438, 983), and (1082, 1387), respectively. The same case also appears in the SN-Q sample. In order to neutralize the inhomogeneous
Pantheon sample, it might be necessary to obtain a larger and more homogeneous sample.

\subsection{The MCMC method}
For the standard cosmological model, the theoretical distance
modulus can be written as
\begin{equation}
\mu_{th} = 5 \log_{10} \frac{d_{L}}{\textnormal{Mpc}} + 25,
\label{q1}
\end{equation}
where $d_{L}$ is the luminosity distance. In the flat $\rm \Lambda$CDM
model, $d_L$ can be calculated from
\begin{equation}
d_{L} = \frac{c(1+z)}{H_{0}} \int_{0}^{z} 
\frac{dz'}{\sqrt{\Omega_{m} (1+z')^{3} + (1-\Omega_{m})}},
\label{q2}
\end{equation}
where $c$ is the speed of light, $H_{0}$ is the Hubble constant, and
$\Omega_{m}$ is the matter density. For SNe Ia, the best fitting
value of $\Omega_{m}$ is achieved by minimizing the value of
$\chi^{2}$
\begin{equation}
        \chi^{2}_{SN} = \sum_{i=1}^{1048}\frac{(\mu_{obs}(z_{i}) - \mu_{th}(\Omega_{m},z_{i}))^{2}}{\sigma^{2}},
\label{q3}
\end{equation}
where $\sigma_{i}(z_{i})$ is the observational uncertainty of
the distance modulus.

For quasars, the relation between UV and X-ray luminosities can be
parameterized as \citep{1986ApJ...305...83A}
\begin{equation}
\log_{10}(L_{X}) = \gamma\log_{10}(L_{UV}) + \beta,
\label{q5}
\end{equation}
where $L_{X}$ is the rest-frame monochromatic luminosity at 2 keV
and $L_{UV}$ is the luminosity at 2,500 {\AA}. We note that $\gamma$ and $\beta$ are
two free parameters. Considering $L = 4\pi d_{L}^{2}F$, 
equation~(\ref{q5}) can be written as
\begin{equation}
\log_{10}(4 \pi d_{L}^{2} F_{X}) = \gamma\log_{10}(4 \pi d_{L}^{2} F_{UV}) + \beta.
\label{q6}
\end{equation}
From equation~(\ref{q6}), we derive the theoretical X-ray flux \citep{2020MNRAS492.4456K},
\begin{eqnarray}
\phi([F_{UV}]_{i},d_{L}[z_{i}]) &=&\log_{10}(F_{X}) \nonumber \\ 
&=& \gamma(\log{F_{UV}}) +(\gamma-1)\log{4\pi} \nonumber\\ 
&+& 2(\gamma-1)\log_{10}{d_{L}} + \beta.
\label{q7}
\end{eqnarray}
For the quasar sample, the form of $\chi^{2}_Q$ related to the X-ray
flux $F_{X}$ of the quasar is given as
\begin{eqnarray}
\chi^{2}_{Q} &=&
\sum_{i=1}^{1421}(\frac{(\log_{10}(F_{X})_{i}-\phi([F_{UV}]_{i},d_{L}[z_{i}]))^{2}}{s_{i}^{2}} \nonumber\\
&+& \ln(2\pi s_{i}^{2})), \label{q4}
\end{eqnarray}
where the variance $s_{i}^{2}$ consists of the global intrinsic
$\delta$ and the measurement error $\sigma_{i}$ in $(F_{X})_{i}$,
that is, $s_{i}^{2} \equiv \delta^{2} + \sigma_{i}^{2}$. The function
$\phi$ corresponds to the theoretical X-ray flux (equation (\ref{q5})). Compared to $\delta$ and $\sigma_{i}^{2}$, the
error of $(F_{UV})_{i}$ is negligible.

In substituting equation (\ref{q7}) for equation (\ref{q4}) and
minimizing the value of $\chi^{2}_{Q}$, we obtain the best-fit
parameters: $\Omega_{m}$ = 0.74$^{+0.22}_{-0.17}$, $\delta$ = 0.23$^{+0.0046}_{-0.0044}$, $\gamma$ =
0.62$^{+0.12}_{-0.10}$, and $\beta$ = 7.53$^{+0.29}_{-0.35}$. It is important to note that $\Omega_{m}$ is in
4$\sigma$ tension with the \textnormal{$\Lambda$CDM model} \citep{2019NatAs...3..272R}. By combining
equations (\ref{q1}), (\ref{q3}), (\ref{q7}), and (\ref{q4}), the $\chi^2$ statistic for the SN-Q sample is
\begin{eqnarray}
\chi^{2}_{Total} &=& \chi^{2}_{SN} + \chi^{2}_Q \nonumber\\ 
&=&\sum_{i=1}^{1048}\frac{(\mu_{obs}(z_{i}) -
                \mu_{th}(\Omega_{m},z_{i}))^{2}}{\sigma^{2}} \nonumber\\ 
&+&\sum_{i=1}^{1421}(\frac{(\log_{10}(F_{X})_{i}-\phi([F_{UV}]_{i},d_{L}[z_{i}]))^{2}}{s_{i}^{2}} \nonumber\\ 
&+& \ln(2\pi s_{i}^{2})). \label{q8}
\end{eqnarray}

\begin{figure*}
        \centering
        \includegraphics[width=1\textwidth]{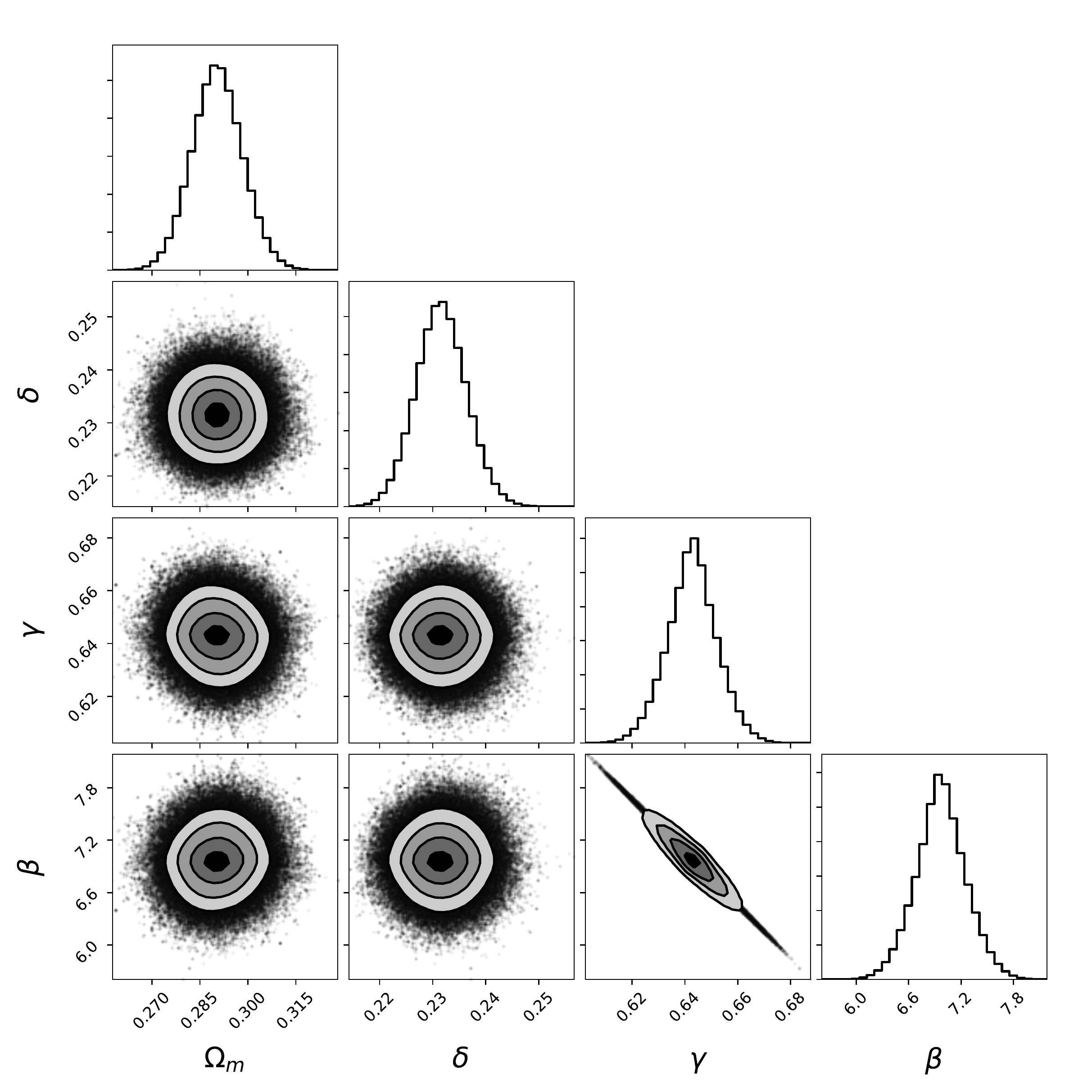}
        \caption{Confidence contours ($1\sigma,2\sigma$, and $3\sigma$) and the marginalized likelihood distributions for the space of the parameters ($\Omega_{m}$, $\delta$, $\gamma$, $\beta$) from the SN-Q sample in the spatially flat $\rm \Lambda$CDM model.}
        \label{F0}       % Give a unique label
\end{figure*}

By using equation (\ref{q8}), the best-fit parameters are
$\Omega_{m}$ = 0.29$^{+0.0075}_{-0.0096}$, $\delta$ = 0.23$^{+0.0046}_{-0.0057}$, $\gamma$ = 0.64$^{+0.0089}_{-0.0115}$, and $\beta$ =
6.98$^{+0.26}_{-0.36}$ from the SN-Q sample, which is shown in
Figure \ref{F0}. The result is consistent with the $\rm \Lambda$CDM model.

From the best-fit parameters for the quasar and SN-Q samples, we
find that when only considering the quasar sample, the matter
density $\Omega_{m}$ = 0.74$^{+0.22}_{-0.17}$. This is a
significant departure from the $\Omega_{m}$ given by the standard
cosmological model. But $\Omega_{m}$ = 0.29$^{+0.0075}_{-0.0096}$ from the SN-Q sample is consistent
with the standard cosmological model. This may imply that the
quasars could not be used for the cosmological probe independently
at present \citep{2020PhRvD101.043502V}. It can be combined with
others probes, for instance, SNe Ia, CMB, GRBs
\citep{Wang15}, and fast radio bursts \citep{Yu17}.

\section{Testing the cosmic anisotropy with the HC and DF methods}

\subsection{The HC method and results}
\label{sec:HC method}
The HC method was first proposed by \citet{2007AA...474..717S};
it is widely used to investigate the cosmic anisotropy. For
example, the anisotropy of cosmic expansion, the temperature
anisotropy of the CMB \citep{2013ApJS..208...20B,2004ApJ...605...14E,2014AA...571A..23P,2014ApJ...784L..42A,2004MNRAS.354..641H,2015JCAP...01..008Q},
and the acceleration scale of modified Newtonian dynamics
\citep{2017ApJ...847...86Z,2018ChPhC..42k5103C,2019MNRAS.486.1658C}. Firstly, we briefly introduce this approach. Its goal is to identify the direction, which corresponds to the axis of maximal asymmetry from the dataset, by comparing the accelerating expansion rate. In the spatially flat $\rm \Lambda$CDM model, it is convenient to employ $\Omega_{m}$ to replace the accelerating expansion rate considering the relationship between the deceleration parameter $q_{0}$ and $\Omega_{m}$. The most important step is to produce random directions $\hat{V}$ $(l$, $b)$, which are used to split the dataset into two parts (defined as "up" and ``down"), where $l\in(0^{\circ}$, $360^{\circ})$ and $b\in(-90^{\circ}$, $90^{\circ})$ are the longitude and latitude in the galactic coordinate system, respectively. According to ``up" and ``down" subdatasets, the corresponding best-fit values of $\Omega_{m,u}$ and $\Omega_{m,d}$ are obtained adopting the MCMC method. The nuisance parameters ($\delta$, $\gamma$ and $\beta$) are marginalized along with $\Omega_{m}$ in the MCMC process. The anisotropy level (AL) made up of $\Omega_{m,u}$ and $\Omega_{m,d}$ can be used to describe the accelerating expansion rate. 

\begin{figure*}
        \centering
        \includegraphics[width=1\textwidth]{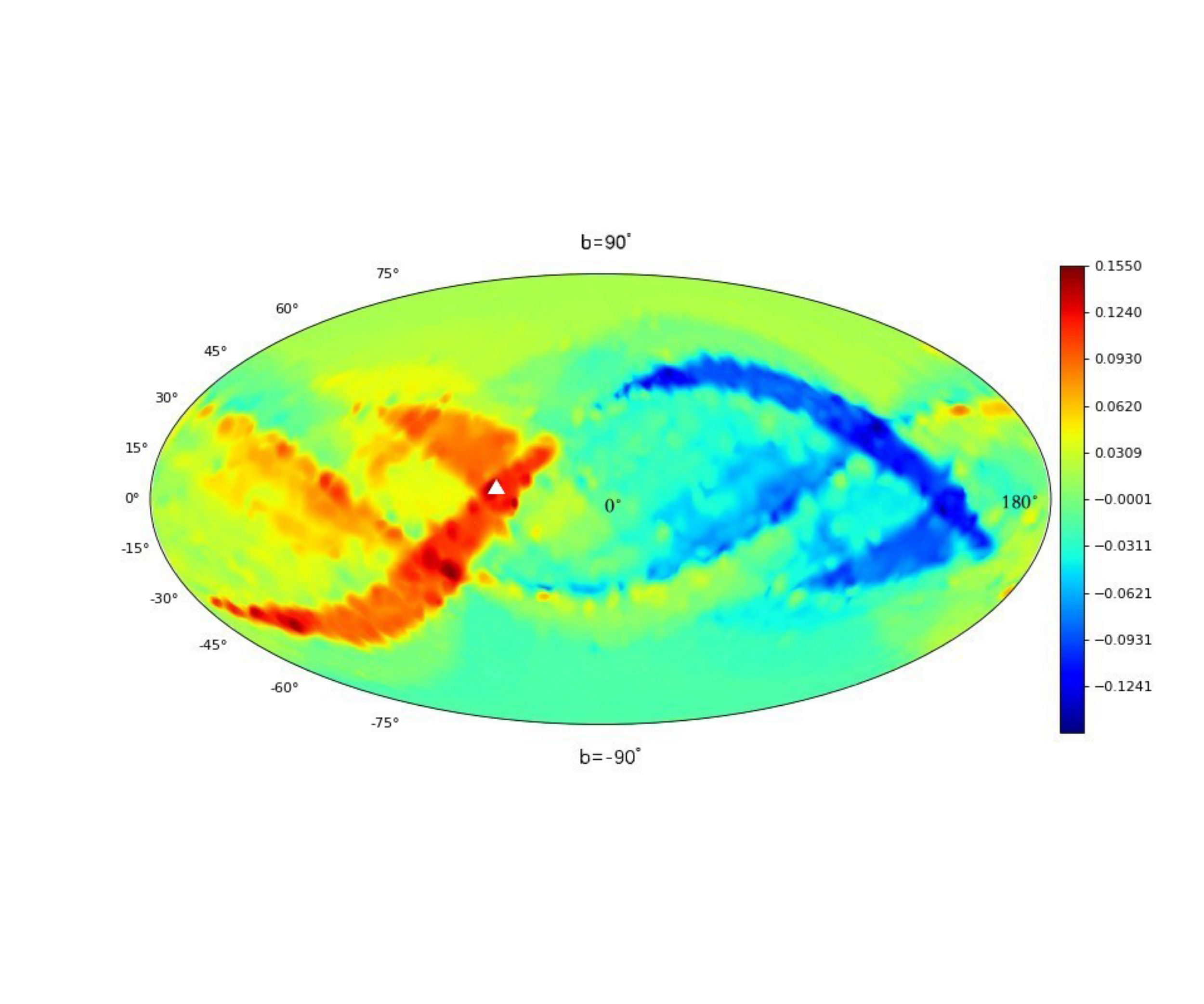}
        \caption{Distribution of the $AL$ in the galactic coordinate system. The triangle marks the direction of the largest of the $AL$ in the sky.}
        \label{F3}       % Give a unique label
\end{figure*}

\begin{figure*}
        \centering
        \includegraphics[width=1\textwidth]{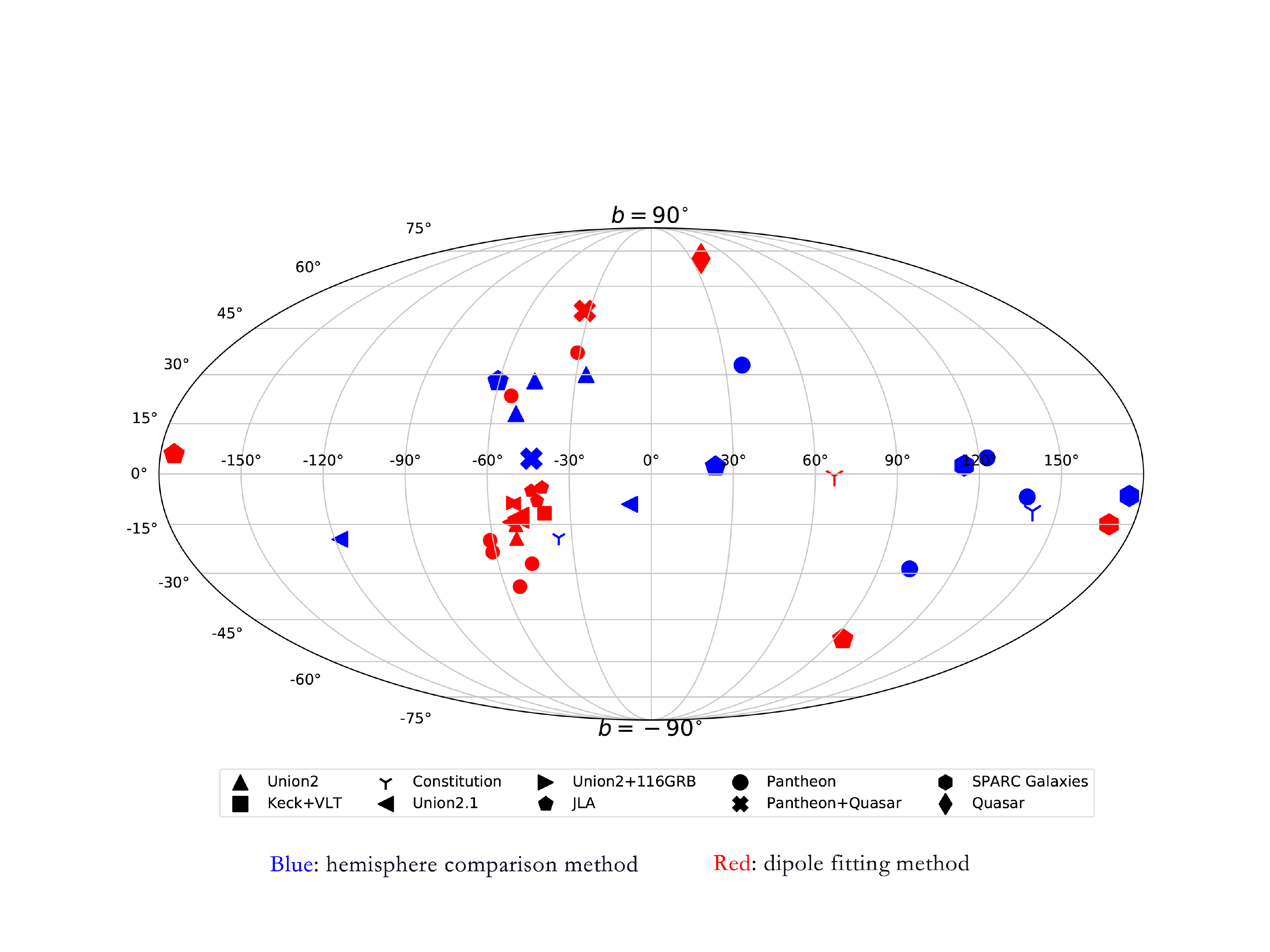}
        \caption{Distribution of the preferred directions ($l$, $b$) in the various observational datasets. The blue color and red color correspond to the HC method and the DF method, respectively.}
        \label{F7}       % Give a unique label
\end{figure*}

In this section, we adopt the HC method and use the SN-Q sample to
study the cosmic anisotropy. The AL is defined as
\begin{equation}
AL = \frac{\triangle \Omega_{m}}{\bar{\Omega}_{m}} = 2 \times \frac{\Omega_{m,u} - \Omega_{m,d}}{\Omega_{m,u} + \Omega_{m,d}},
\label{q9}
\end{equation}
where $\Omega_{m,u}$ and $\Omega_{m,d}$ are the best-fit
$\Omega_{m}$ of the ``up" subset and ``down" subset, respectively. These
two subsets are distinguished from the full SN-Q sample by a random
direction $\hat{V}$$(l,b)$.
The $1\sigma$ uncertainty $\sigma_{AL}$ is
\begin{equation}
\sigma_{AL} = \frac{\sqrt{\sigma^{2}_{\Omega^{\textnormal{max}}_{m,u}} + \sigma^{2}_{\Omega^{\textnormal{max}}_{m,d}}}}{\Omega^{\textnormal{max}}_{m,u} + \Omega^{\textnormal{max}}_{m,d}}.
\label{q10}
\end{equation}
During the calculation, we repeated 3000 random directions $\hat{V}$$(l,b)$
for the HC method and the results are
$\Omega^{\textnormal{max}}_{m,u}$ = 0.407 and
$\Omega^{\textnormal{max}}_{m,d}$ = 0.422, and the corresponding
errors are $\sigma^{2}_{\Omega^{\textnormal{max}}_{m,d}}$ = 0.0149
and $\sigma^{2}_{\Omega^{\textnormal{max}}_{m,d}}$ = 0.0156. Then, we obtain the $1\sigma$ uncertainty $\sigma_{AL} = 0.026$. The
anisotropic level with $1\sigma$ uncertainty is
\begin{equation}
AL_{\textnormal{max}} = 0.142 \pm 0.026,
\label{q11}
\end{equation}
and the corresponding direction is
\begin{equation}
(l, b) = ({316.08^{\circ}}^{+27.41}_{-129.48}, \ {4.53^{\circ}}^{+26.29}_{-64.06}).
\label{q12}
\end{equation}
The distribution of AL$(l$, $b)$ is shown in Figure \ref{F3}. This preferred direction is inconsistent with the previous results with the Pantheon sample from the HC method. For instance, by employing the HC method, \cite{2018PhRvD..97l3515D} obtained two preferred directions (138.8$^{\circ}$, $-$6.8$^{\circ}$) and (102.36$^{\circ}$, $-$28 .58$^{\circ}$) from the Pantheon sample. In using the same sample, \cite{2018MNRAS.478.5153S} and \cite{2019MNRAS.486.5679Z} achieved the HC preferred direction ($l$, $b$) = (37$^{\circ}$, 33$^{\circ}$) and ($l$, $b$) = (123.05$^{\circ}$, 4.78$^{\circ}$), respectively. These four directions are plotted in Figure \ref{F7}. From the result of the HC method, it shows that the quasar sample has an obvious impact on the HC results. 

\begin{figure*}
        \centering
        \includegraphics[width=0.6\textwidth]{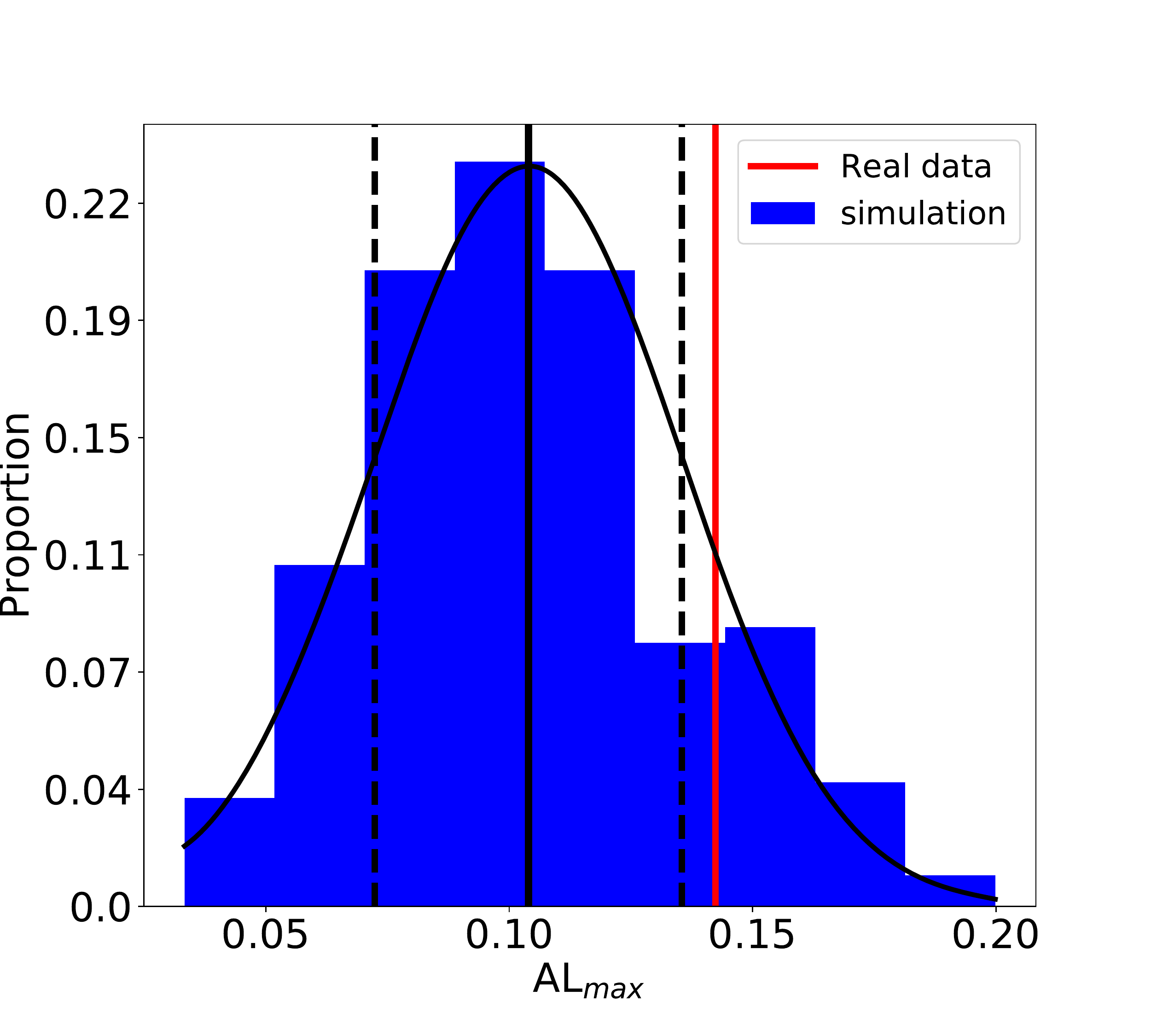}
        \caption{Distribution of AL$_{max}$ in 200 simulated datasets. The black curve is the best fitting result to the Gaussian function. The solid black and vertical dashed lines are commensurate with the mean and the standard deviation, respectively. The vertical red line shows the maximum AL derived from the actual SN-Q sample. }
        \label{random}       % Give a unique label
\end{figure*}

\begin{figure*}
        \centering
        \includegraphics[width=0.6\textwidth]{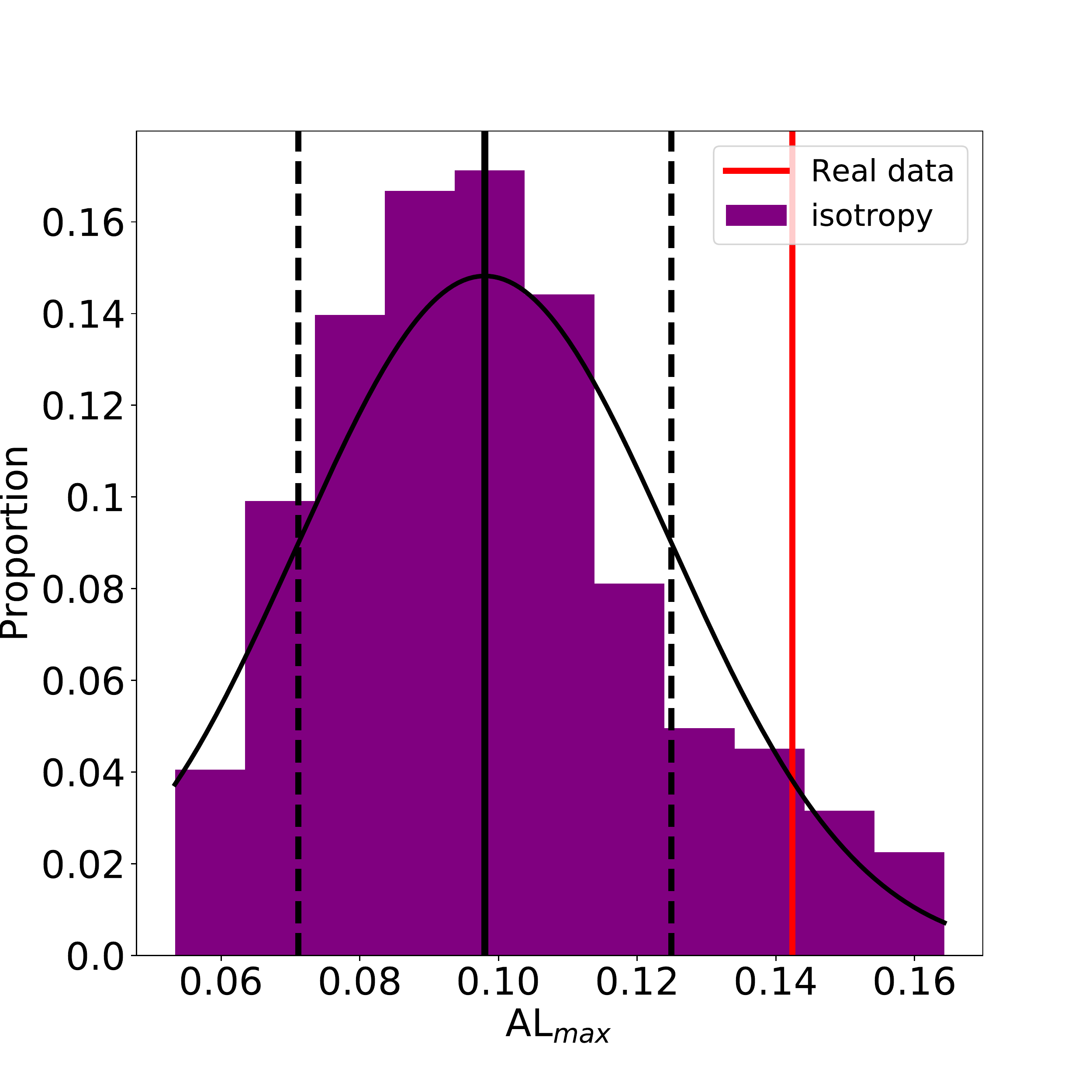}
        \caption{Distribution of AL$_{max}$ in 200 simulated isotropic datasets. The black curve is the best fitting result to the Gaussian function. The solid black and vertical dashed lines are commensurate with the mean and the standard deviation, respectively. The vertical red line shows the maximum AL derived from the actual SN-Q sample. }
        \label{iso}       % Give a unique label
\end{figure*}

We also assessed statistical significance of the results found by means of this test with simulated datasets. The simulated SN-Q datasets have the same direction in the sky as real data, but a different distance modulus (SNe Ia) and X-ray flux (quasar). Then the corresponding distance modulus and X-ray flux were constructed by the Gaussian function with the mean determined by equation (\ref{q8}), where $\Omega_{m}$ = 0.29, $\delta$ = 0.23, $\gamma$ = 0.64, $\beta$ = 6.98 is the best-fitting value to the SN-Q sample and the standard deviation equal to the corresponding real datapoint 1$\sigma$ error. Therefore, we constructed 200 simulated datasets and obtain the corresponding AL$_{max}$ by employing the HC method. The distribution of AL$_{max}$ in these datasets was fitted by a Gaussian function with the mean value 0.104 and the standard deviation 0.032, as shown in Figure \ref{random}. From Figure \ref{random}, we can find that the statistical significance of the maximum anisotropy level of AL$_{max}$, which is about 1.23$\sigma$ and 14 percent of  AL$_{max}$ obtained from the simulated datasets, is greater than that of the real data. Therefore, the value of AL$_{max}$ is not large enough. In addition, it is necessary to examine whether the maximum AL from the SN-Q sample is consistent with statistical isotropy. In order to determine this, we evenly redistributed the original data-sets across the sky and repeated it 200 times. The result of the simulated isotropic datasets is shown in Figure \ref{iso}. The mean value is 0.099 and the standard deviation is 0.024. We note that 7 percent of  AL$_{max}$ obtained from the simulated isotropic datasets is larger than that of the real data. The statistical significance of AL$_{max}$ is about 1.65$\sigma$. The results of the random and isotropic datasets are smaller than 2$\sigma$, which still supports the absence of significant anisotropy.

In the end of this subsection, we compare the HC preferred direction of the SN-Q sample with those derived in various observational datasets. In Figure \ref{F7}, we plotted the preferred directions ($l$, $b$) found from various observational datasets by the HC method in the galactic coordinate system. From Figure \ref{F7}, we find that the HC preferred direction in this paper is the deviation from that of the Pantheon sample and the SPARC galaxies sample, but it is generally consistent with those in the Union2 sample \citep{2010JCAP...12..012A,2012JCAP...02..004C,2015MNRAS.446.2952C}, the Union2.1 sample \citep{2019EPJC...79..783S,2016MNRAS.460..617L}, the Constitution sample \citep{2019EPJC...79..783S,2013AA...553A..56K}, and the JLA sample \citep{2018PhRvD..97l3515D}. Overall, the distribution of the HC preferred direction obtained from a different sample is diffuse. As discussed by \cite{2019EPJC...79..783S} and \cite{2019MNRAS.486.5679Z}, the cosmic anisotropy found in the supernova sample
significantly relies on the inhomogeneous distribution of SNe Ia in
the sky. The preferred directions in various observational datasets are also shown in Table \ref{T2}. From Table \ref{T2}, we find that the HC preferred direction in this paper is also in agreement with the results of the CMB dipole \citep{1996ApJ...470...38L}, velocity flows \citep{2009MNRAS.392..743W,2010MNRAS.407.2328F}, quasar alignment \citep{2005AA...441..915H,2011ASPC..449..441H,2001AA...367..381H}, the CMB quadrupole \citep{2004MNRAS.355.1283B,2010MNRAS.403.1739F}, the CMB octopole \citep{2004MNRAS.355.1283B},  $\bigtriangleup$$\alpha$/$\alpha$ \citep{2011PhRvL.107s1101W,2012MNRAS.422.3370K}, and infrared galaxies \citep{2014MNRAS.445.L60Y,2017MNRAS...464...768B}.

\subsection{The DF method and results}
The DF method is also used to test the cosmic anisotropy. Considering
the dipole magnitude $A$ and monopole term $B$, the theoretical distance
modulus should be rewritten as
\begin{equation}
\tilde{\mu}_{th} = \mu_{th}\times(1+A(\hat{\textbf{n}}\cdot\hat{\textbf{p}}) + B),
\label{q13}
\end{equation}
where $\hat{\textbf{n}}$ and $\hat{\textbf{p}}$ correspond to the
dipole direction and the unit vector pointing to the position of the SN Ia or quasar, respectively. In
the galactic coordinate, $\hat{\textbf{n}}$ is written as
\begin{equation}
\hat{\textbf{n}} = \cos{(b)}\cos{(l)}\hat{\textbf{i}}+\cos{(b)}\sin{(l)}\hat{\textbf{(j)}}+sin{(b)}\hat{\textbf{k}}.
\label{q14}
\end{equation}
For any $i_{th}(l_{i},b_{i})$ data points, $\hat{\textbf{p}}$ is given by
\begin{equation}
\hat{\textbf{p}_{i}} = \cos{(b_{i})}\cos{(l_{i})}\hat{\textbf{i}}+\cos{(b_{i})}\sin{(l_{i})}\hat{\textbf{(j)}}+sin{(b_{i})}\hat{\textbf{k}}.
\label{q15}
\end{equation}

\begin{figure*}
        \centering
        \includegraphics[width=1\textwidth]{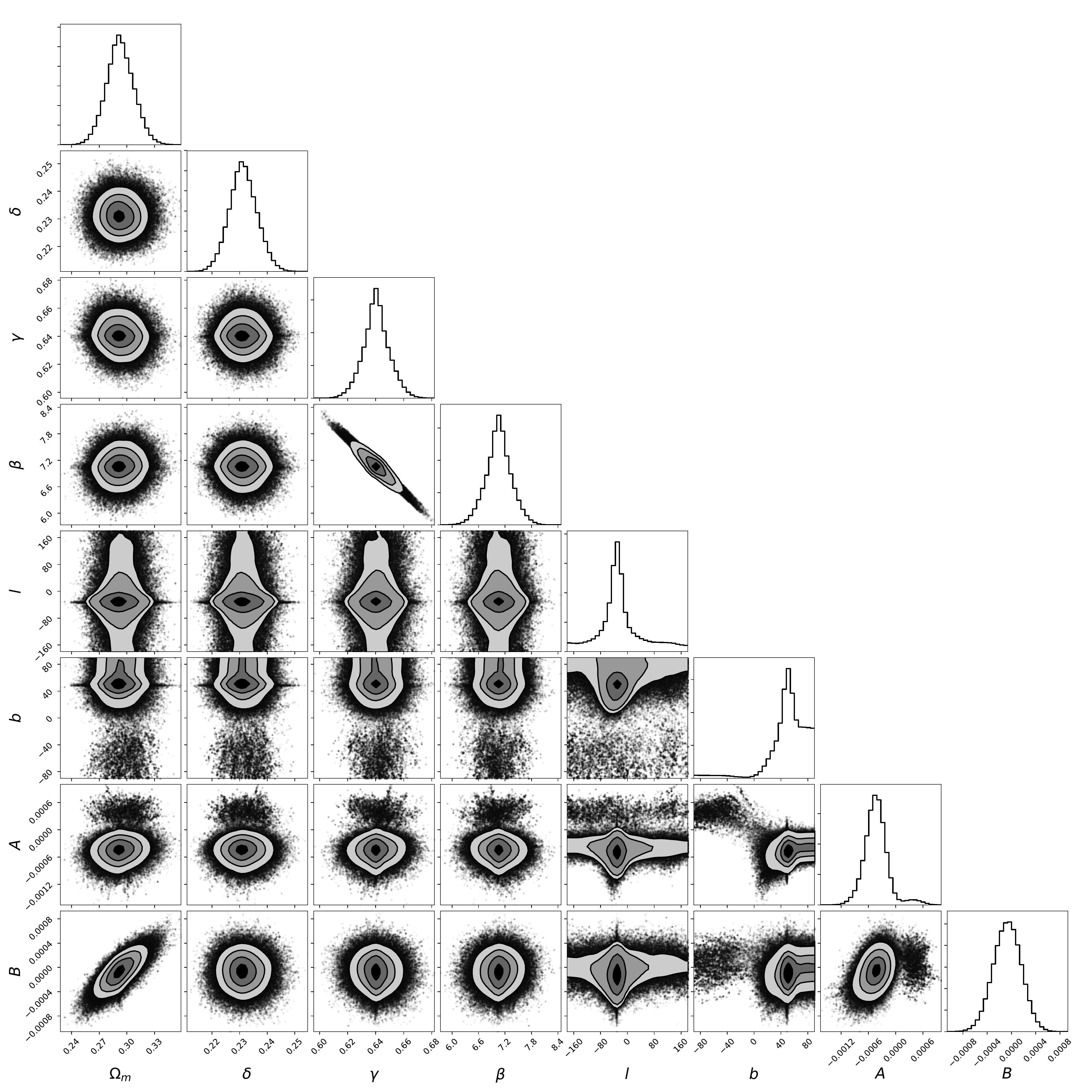}
        \caption{Confidence contours ($1\sigma,2\sigma,$ and $3\sigma$) and marginalized likelihood distributions for the
                parameters space ($\Omega_{m}$, $\delta$, $\gamma$, $\beta$, $l$, $b$, $A$, $B$) to the SN-Q sample in the dipole-modulated $\rm \Lambda$CDM model.}
        \label{F5}       % Give a unique label
\end{figure*}

We were able to obtain the best-fit dipole direction ($l$, $b$) by substituting
equation (\ref{q13}) for equation (\ref{q3}) and minimizing
$\chi^{2}$. Equation (\ref{q13}) can be directly used for the SNe Ia
sample. For the quasars, according to equation (\ref{q1}), the theoretical distance
modulus can also be written as
\begin{eqnarray}
\tilde{\mu}_{th} = 5 \tilde{\log}_{10} \frac{d_{L}}{\textnormal{Mpc}} + 25.
\label{q17}
\end{eqnarray}
In substituting equation \ref{q17} for equation \ref{q13}, we obtain
\begin{equation}
\tilde{\log}_{10} \frac{d_{L}}{\textnormal{Mpc}} = (\log_{10} \frac{d_{L}}{\textnormal{Mpc}} + 5) \times(1+A(\hat{\textbf{n}}\cdot\hat{\textbf{p}}) + B) - 5.
\label{q18}
\end{equation}
Thus the theoretical X-ray flux with dipole and monopole corrections is given by
\begin{eqnarray}
\tilde{\phi}([F_{UV}]_{i}, d_{L}[z_{i}]) &=& \gamma(\log{F_{UV}}) +(\gamma-1)\log{4\pi} + \beta \nonumber\\
&+& 2(\gamma-1)((\log_{10} \frac{d_{L}}{\textnormal{Mpc}} + 5) \nonumber\\
&\times& (1+A(\hat{\textbf{n}}\cdot\hat{\textbf{p}}) + B) - 5).
\label{q19}
\end{eqnarray}
Adding equations
(\ref{q13}) and (\ref{q19}) to (\ref{q8}) and minimizing the value of
$\chi^{2}_{Total}$, the marginalized posterior distribution for the
SN-Q sample is shown in Figure \ref{F5}. In total, there are eight
parameters in the DF method, four of which are used to describe the
universal anisotropy, that is, $l$, $b$, $A,$ and $B$. The results show
that the dipole direction ($l$, $b$) is ($327.21^{\circ}$$^{+28.24}_{-27.50}$, $48.67^{\circ}$$^{+8.99}_{-8.54}$), the dipole magnitude $A$ = ($-$4.36$^{+1.98}_{-2.74}$)$\times$$10^{-4}$ and the monopole term $B$ = ($-$6.60$^{+21.5}_{-28.6}$)$\times$$10^{-5}$. The dipole direction is generally consistent with the
preferred direction $({316.08^{\circ}}^{+27.41}_{-129.48}, \ {4.53^{\circ}}^{+26.29}_{-64.06})$ of the HC method. Both the dipole magnitude $A$ and monopole term $B$ are approximately equal to zero, indicating that there is no significant anisotropy in the SN-Q sample. Next, we carry out a redshift tomography analysis to discuss the effect of redshift. The results of the redshift tomography analysis were calculated, as shown in Table \ref{T1}. Based on analyses of Figure \ref{F1}, the
number of low redshift sources is relatively large, so the redshift intervals are not uniform in the redshift tomography analysis. From the results of redshift tomography, it can be found that the dipole directions are distributed in a relatively small range. The maximum of monopole term $A$ and the dipole magnitude $B$ are near zero. There is no significant change in the dipole direction and anisotropic level with different redshift ranges. We note that |$A$| and |$B$| of the first bin are larger than that of others bins. This indicates that the anisotropy level of the low redshift range might be relatively higher.

At the end of this section, we make a comparison between the dipole directions of the SN-Q sample with that derived from other samples and compare the results between the HC method and the DF method for the same sample. At first, we marked the dipole directions obtained from different samples in the galactic coordinate system, as shown in Figure \ref{F7}. The dipole directions obtained from different samples are mostly located in a relatively small part of the South Galactic Hemisphere. The dipole direction in this paper is inconsistent with them. The longitude $l$ is close, but the deviation of latitude $b$ is large. It might be caused by the inhomogeneous distribution of SNe Ia, as discussed in Section \ref{sec:sample}. The data number of the SN-Q sample in the North Galactic Hemisphere is larger than that in the South Galactic Hemisphere. But the dipole direction is in agreement with the results derived for the CMB dipole, velocity flows, quasar alignment, the CMB quadrupole, the CMB octopole, and $\bigtriangleup$$\alpha$/$\alpha$. For the same sample, we find that the results of these two methods are not always consistent with each other. For instance, these two preferred directions obtained by the HC and DF methods in the SPARC galaxies sample \citep{2017ApJ...847...86Z,2018ChPhC..42k5103C}, the Union2 sample \citep{2010JCAP...12..012A,2012PhRvD..86h3517M,2012JCAP...02..004C,2013PhRvD..87l3522C,2015MNRAS.446.2952C}, and the JLA sample \citep{2015ApJ...808...39B,2019EPJC...79..783S,2016MNRAS.456.1881L,2018PhRvD..97l3515D}, are consistent. However, the HC and DF preferred directions obtained from the Constitution sample \citep{2019EPJC...79..783S,2013AA...553A..56K} are inconsistent. The same case is also exhibited in the Pantheon sample \citep{2018MNRAS.478.5153S,2019MNRAS.486.5679Z,2018EPJC...78..755D}. This may be due to the sensitivity of the two methods \citep{2015MNRAS.446.2952C}.  

In Table \ref{T2}, we summarize the preferred directions ($l$, $b$) found in different cosmological models using different methods and observational datasets. The parts of the DF method and HC method have been discussed at the beginning of this section. In addition, we also summarize the research results obtained by various methods for different models ($\rm \Lambda$CDM, $w$CDM, and CPL). Looking through the results from the same sample, we find that the results are almost independent of these three models.

\section{Summary}

The cosmic principle assumes that the universe is homogeneous and
isotropic on cosmic scales. The research on cosmic anisotropy from SNe Ia also
shows that there is no obvious anisotropy. In this work, we test the
cosmic anisotropy with a new sample, which consists of SNe
Ia and the quasars, by using the HC and DF methods.
Nevertheless, the results show that there is no obvious
anisotropy.

Firstly, we briefly investigate the Pantheon sample and quasar sample. By assessing the results, we find that adding the quasar sample reduces the unevenness of the overall sample. Compared with the Pantheon sample, the new sample (SN-Q sample) has a larger size, wider range of redshift, and a more uniform distribution. But the distribution of the SN-Q sample is still inhomogeneous in the North Galactic Hemisphere and South Galactic Hemisphere. We also briefly discuss the effect of inhomogeneous distribution on the dipole preferred in the SN-Q sample. By adopting the MCMC method, we obtain the best fitting values of $\Omega_{m}$, $\delta$, $\gamma$, $\beta$, and $H_{0}$ and find that 4$\sigma$ tensions exist between the best-fit value of $\Omega_{m}$ from the quasar sample and that of the $\rm \Lambda$CDM model. Then, by adding the Pantheon sample into the quasar sample, the 4$\sigma$ tension disappears.

For the HC method, the preferred direction that corresponds to the
maximum accelerating expansion is 
${316.08^{\circ}}^{+27.41}_{-129.48}$, ${4.53^{\circ}}^{+26.51}_{-64.06}$. The anisotropy level $AL_{\textnormal{max}}$ is equal to 0.142$\pm$ 0.026. By using the simulated datasets, we assess the HC result and determine that the statistical significance is about 1.23$\sigma$. In addition, we also examine whether the $AL_{\textnormal{max}}$ from the SN-Q sample is consistent with statistical isotropy by employing the simulated isotropic datasets. The statistical significance is about 1.65$\sigma$. The results show that it is hardly reproduced by simulated datasets or isotropy simulated datastes. For the DF method, we find that the preferred direction in the SN-Q sample points toward ($327.21^{\circ}$$^{+28.24}_{-27.50}$, $48.67^{\circ}$$^{+8.99}_{-8.54}$) with an anisotropy level of $A$ = ($-$4.36$^{+1.98}_{-2.74}$) $\times$ $10^{-4}$, which is marginally consistent with the result of the HC method. There is a considerable deviation for the latitude direction $l$ from those obtained from other SNe Ia samples, which may be caused by including the quasar sample. The preferred directions obtained by using the HC method and the DF method from the SN-Q sample are both in agreement with the results of the CMB dipole, velocity flows, quasar alignment, the CMB quadrupole, the CMB octopole, and  $\bigtriangleup$$\alpha$/$\alpha$. The results of the redshift tomographic analysis show that the dipole direction is weakly dependent on redshift. Comparing this with previous studies of the Pantheon sample, the preferred directions in the SN-Q sample have an obvious divergence. There does not exist an obvious anisotropy from the SN-Q sample.

Although the SN-Q sample is better than the Pantheon sample in some aspects, such as the quantity, redshift range, and uniform of distribution, there are still some shortcomings. For example, the distribution of the SN-Q sample is not uniform in the whole sky. Next-generation X-ray surveys, such as eROSITA, will provide us with larger and more precise luminosity distance determinations of quasars, so that we should be able to reduce the uncertainties obtained in our analysis. By employing a simulated e-ROSITA quasar sample, predictions for e-ROSITA from a quasar Hubble diagram have been made by \citet{2020FrASS...7....8L}. \citet{2019AA.631.L13C} point out that the cosmic acceleration is due to a non-negligible dipole anisotropy by analyzing the JLA sample. We think that the same research can be performed by a combined sample of SNe Ia and quasars, considering the quasar sample enables us to talk about this study in a higher redshift range and to test whether the same result will still be obtained. This will be pursued in future work.

\begin{acknowledgements}
We thank the anonymous referee for valuable comments. This work is supported by the National Natural Science Foundation of
China (grant U1831207).
\end{acknowledgements}


\onecolumn
%%%%%%%%%%%%%%%%%%%%%%%%%%%%%%%%%%%%%%%%%%%%%%%%%%
%Fig.1

%%%%%%%%%%%%%%%%%%%%%%%%%%%%%%%%%%%%%%%%%%%%%%%%%%
%Fig.2

%%%%%%%%%%%%%%%%%%%%%%%%%%%%%%%%%%%%%%%%%%%%%%%%%%
%Fig.3

%%%%%%%%%%%%%%%%%%%%%%%%%%%%%%%%%%%%%%%%%%%%%%%%%%
%Fig.4

%%%%%%%%%%%%%%%%%%%%%%%%%%%%%%%%%%%%%%%%%%%%%%%%%%
%Fig.5

%%%%%%%%%%%%%%%%%%%%%%%%%%%%%%%%%%%%%%%%%%%%%%%%%%
%Fig.6

%%%%%%%%%%%%%%%%%%%%%%%%%%%%%%%%%%%%%%%%%%%%%%%%%%
%Fig.7

%%%%%%%%%%%%%%%%%%%%%%%%%%%%%%%%%%%%%%%%%%%%%%%%%%
%Fig.8

%%%%%%%%%%%%%%%%%%%%%%%%%%%%%%%%%%%%%%%%%%%%%%%%%%
%Tab.1

\begin{table}
        \centering
        \caption{ Preferred directions ($l$, $b$) found in different cosmological models using different methods and observational datasets.}
        \label{T2}       % Give a unique label
        \begin{spacing}{1.2}
                \begin{tabular}{|p{3.15cm}|p{1.5cm}|p{1.8cm}|p{2.5cm}|p{2.5cm}|p{3.5cm}|}
                        \hline
                        \hline
                        Cosmological Obs.& Model & Method &  $l$ ($^{\circ})$ & $b$ ($^{\circ})$ & Ref.  \\ \hline
                        Union2 & $\rm \Lambda$CDM & HC & 309$^{\circ}$$^{+23}_{-3}$ & 18$^{\circ}$$^{+11}_{-10}$  & \citet{2010JCAP...12..012A} \\ 
                        & $\rm \Lambda$CDM & - & 309$^{\circ}$ & 19$^{\circ}$  & \citet{2011MNRAS.414.264C}  \\                       
                        & $\rm \Lambda$CDM & HC & 314$^{\circ}$$^{+20}_{-13}$ & 28$^{\circ}$$^{+11}_{-33}$  & \citet{2012JCAP...02..004C}  \\ 
                        & $\rm \Lambda$CDM & DF & 309.4$^{\circ}$$\pm$18.0   & -15.1$^{\circ}$$\pm$11.5    & \citet{2012PhRvD..86h3517M}  \\ 
                        & $\rm \Lambda$CDM & HC & 334$^{\circ}$$^{+6}_{-6}$   & 30$^{\circ}$$^{+2}_{-8}$      & \citet{2015MNRAS.446.2952C}  \\ 
                        & $\rm \Lambda$CDM & DF & 309$^{\circ}$$\pm$22.4     & -19.3$^{\circ}$$\pm$12.9        & \citet{2015MNRAS.446.2952C}  \\ 
                        & $\rm \Lambda$CDM & AM & 126$^{\circ}$$^{+17}_{-26}$ & 13$^{\circ}$$^{+19}_{-25}$ & \citet{2013PhRvD..87l3522C}  \\ \hline
                        67GRB & $\rm \Lambda$CDM & AM & 336$^{\circ}$$^{+33}_{-223}$ & -5$^{\circ}$$^{+34}_{-26}$ &  \citet{2013PhRvD..87l3522C}  \\ 
                        & $w$CDM  & AM & 340$^{\circ}$$^{+35}_{-227}$ & -4$^{\circ}$$^{+34}_{-30}$ & \citet{2013PhRvD..87l3522C}  \\
                        & CPL          & AM & 339$^{\circ}$$^{+26}_{-224}$ & -6$^{\circ}$$^{+34}_{-25}$ & \citet{2013PhRvD..87l3522C}  \\ \hline
                        Union2+67GRB & $\rm \Lambda$CDM & AM & 129$^{\circ}$$^{+16}_{-23}$    & 16$^{\circ}$$^{+17}_{-10}$ &          \citep{2013PhRvD..87l3522C}  \\ 
                        & $w$CDM  & AM & 129$^{\circ}$$^{+17}_{-21}$  & 15$^{\circ}$$^{+18}_{-11}$ & \citet{2013PhRvD..87l3522C}  \\ 
                        & CPL          & AM & 131$^{\circ}$$^{+14}_{-22}$  & 16$^{\circ}$$^{+15}_{-10}$ & \citet{2013PhRvD..87l3522C} \\ \hline        
                        
                        Union2 + $\rm SN_{FACTORY}$  & $\rm \Lambda$CDM & DF (z,0.015-0.035) & 298$^{\circ}$$\pm$25 & 15$^{\circ}$$\pm$20 & \citet{2013AA.560.A90F} \\ 
                        & $\rm \Lambda$CDM & DF (z,0.035-0.045) & 302$^{\circ}$$\pm$48 & -12$^{\circ}$$\pm$26 & \citet{2013AA.560.A90F} \\ 
                        & $\rm \Lambda$CDM & DF (z,0.045-0.060) & 359$^{\circ}$$\pm$32 & 14$^{\circ}$$\pm$27 & \citet{2013AA.560.A90F} \\ 
                        & $\rm \Lambda$CDM & DF (z,0.060-0.100) & 285$^{\circ}$$\pm$234 & -23$^{\circ}$$\pm$112 & \citet{2013AA.560.A90F} \\ \hline 
                        
                        Union2.1  & $\rm \Lambda$CDM & HC & -- & -- & \citet{2014MNRAS.437.1840Y}  \\ 
                        & $\rm \Lambda$CDM & DF & 307.1$^{\circ}$$\pm$16.2     & -14.3$^{\circ}$$\pm$10.1    & \citet{2014MNRAS.437.1840Y}  \\ 
                        & $\rm \Lambda$CDM & HC & 241.9$^{\circ}$   & -19.5$^{\circ}$  &  \citep{2016MNRAS.460..617L} \\ 
                        & $\rm \Lambda$CDM & DF & 310.6$^{\circ}$$\pm$18.2   & -13.0$^{\circ}$$\pm$11.1    & \citet{2016MNRAS.460..617L}  \\ 
                        
                        & $\rm \Lambda$CDM & Hubble map & 326.25$^{\circ}$  & 12.02$^{\circ}$  & \citet{2015ApJ...808...39B}  \\ 
                        & $\rm \Lambda$CDM & $q$ map    & 354.38$^{\circ}$  & 27.28$^{\circ}$  & \citet{2015ApJ...808...39B}  \\ 
                        
                        & $\rm \Lambda$CDM & HC & 352$^{\circ}$  &  -9$^{\circ}$ & \citet{2019EPJC...79..783S}  \\ 
                        & $\rm \Lambda$CDM & DF & 309.3$^{\circ}$$^{+15.5}_{-15.7}$  &  -8.9$^{\circ}$$^{+11.2}_{-9.8}$      & \citet{2019EPJC...79..783S}  \\ \hline
                        Union2.1+116GRB  & $\rm \Lambda$CDM & DF & 309.2$^{\circ}$$\pm$15.8 & -8.6$^{\circ}$$\pm$10.5 & \citet{2014MNRAS.443.1680W} \\ \hline

                        Keck+VLT & $\rm \Lambda$CDM & DF & 320.5$^{\circ}$$\pm$11.8 & -11.7$^{\circ}$$\pm$7.5 & \citet{2012PhRvD..86h3517M}  \\ \hline
                        
                        Constitution & $\rm \Lambda$CDM & HC &  -35$^{\circ}$  &  -19$^{\circ}$  & \citet{2013AA...553A..56K}  \\ 
                        & $\rm \Lambda$CDM & HC &    141$^{\circ}$  &   -11$^{\circ}$  & \citet{2019EPJC...79..783S}  \\ 
                        & $\rm \Lambda$CDM & DF & 67.0$^{\circ}$$^{+66.5}_{-66.2}$    & -0.6$^{\circ}$$^{+25.2}_{-26.3}$    & \citet{2019EPJC...79..783S}  \\ \hline
                        
                        JLA  & $\rm \Lambda$CDM & Hubble map & 58.00$^{\circ}$ & -60.43$^{\circ}$  & \citet{2015ApJ...808...39B}  \\ 
                        & $\rm \Lambda$CDM & $q$ map    & 225.00$^{\circ}$ & 51.26$^{\circ}$  & \citet{2015ApJ...808...39B}  \\ 
                        
                        & $\rm \Lambda$CDM & HC(max)    &  23.49$^{\circ}$ & 2.25$^{\circ}$   & \citet{2018PhRvD..97l3515D}  \\ 
                        & $\rm \Lambda$CDM & HC(submax) &  299.47$^{\circ}$ & 28.39$^{\circ}$ & \citet{2018PhRvD..97l3515D}  \\ 
                        & $\rm \Lambda$CDM & DF &  185$^{\circ}$$^{175}_{-185}$ & 5.9$^{\circ}$$^{+84.1}_{-95.9}$ & \citet{2018PhRvD..97l3515D}  \\ 
                        
                        & $\rm \Lambda$CDM & DF    &  316$^{\circ}$$^{+107}_{-100}$    & -5$^{\circ}$$^{+41}_{-60}$    & \citet{2016MNRAS.456.1881L}  \\ 
                        & $w$CDM  & DF    &  320$^{\circ}$$^{+107}_{-104}$     & -4$^{\circ}$$^{+45}_{-61}$    & \citet{2016MNRAS.456.1881L}  \\ 
                        
                        \hline
                        
                \end{tabular}
        \end{spacing}
\end{table}

\begin{table}
        
        \centering
        \begin{spacing}{1.2}
                \begin{supertabular}{|p{3.15cm}|p{1.5cm}|p{1.8cm}|p{2.5cm}|p{2.5cm}|p{3.5cm}|} % four columns, alignment for each
                        \hline
                        & CPL          & DF    &  318$^{\circ}$$^{+177}_{-183}$    & -8$^{\circ}$$^{+36}_{-54}$ & \citet{2016MNRAS.456.1881L}  \\ 
                        & $\rm \Lambda$CDM & DF &    94.4$^{\circ}$    &   -51.7$^{\circ}$    & \citet{2019EPJC...79..783S}  \\ \hline
                        
                        Pantheon & $\rm \Lambda$CDM & HC &  37$^{\circ}$$\pm$40          &  33$^{\circ}$ $\pm$ 16         & \citet{2018MNRAS.478.5153S}  \\ 
                        & $\rm \Lambda$CDM & DF & 329$^{\circ}$$^{+101}_{-28}$  & 37$^{\circ}$$^{+52}_{-21}$     & \citet{2018MNRAS.478.5153S}  \\ 
                        & $\rm \Lambda$CDM & HC(max) & 138.08$^{\circ}$$^{+3.16}_{-16.90}$  & -6.8$^{\circ}$$^{+13.55}_{-2.31}$  & \citet{2018EPJC...78..755D}  \\ 
                        & $\rm \Lambda$CDM & HC(submax) & 102.36$^{\circ}$$^{+47.95}_{-34.22}$  & -28.58$^{\circ}$$^{+50.60}_{-1.78}$ & \citet{2018EPJC...78..755D}  \\ 
                        & $\rm \Lambda$CDM & DF         &  -    &  -     & \citet{2018EPJC...78..755D} \\ 
                        & $\rm \Lambda$CDM & HC &  123.05$^{\circ}$$^{+11.25}_{-4.22}$ &   4.78$^{\circ}$$^{+1.80}_{-8.36}$   & \citet{2019MNRAS.486.5679Z} \\ 
                        & $\rm \Lambda$CDM & DF & 306.00$^{\circ}$$^{+82.95}_{-125.01}$  & -34.20$^{\circ}$$^{+16.82}_{-54.93}$   & \citet{2019MNRAS.486.5679Z}  \\ 
                        & $\rm \Lambda$CDM & HC & 286.93$^{\circ}$$\pm$18.52$^{\circ}$  & 27.02$^{\circ}$$\pm$6.5$^{\circ}$   & \citet{2020PhRvD102.023520K}  \\    
                        & $\rm \Lambda$CDM & DF & 210.25$^{\circ}$$\pm$136.56$^{\circ}$  & 72.85$^{\circ}$$\pm$60.63$^{\circ}$   & \citet{2020PhRvD102.023520K}  \\      
                        & $\rm \Lambda$CDM & DF &  306.00$^{\circ}$$^{+91.94}_{-125.98}$ &   -23.41$^{\circ}$$^{+22.97}_{-54.71}$   & \citet{2019MNRAS.486.1658C} \\
                        & $w$CDM  & DF & 298.81$^{\circ}$$^{+84.18}_{-118.71}$  & -19.80$^{\circ}$$^{+14.07}_{-63.25}$   & \citet{2019MNRAS.486.1658C}  \\ 
                        &     CPL      & DF &  313.20$^{\circ}$$^{+75.30}_{-133.15}$ &   -27.00$^{\circ}$$^{+18.72}_{-57.24}$   & \citet{2019MNRAS.486.1658C} \\ 
                        & Finslerian   & DF & 298.80$^{\circ}$$^{+75.31}_{-118.69}$  & -23.41$^{\circ}$$^{+19.26}_{-57.41}$   & \citet{2019MNRAS.486.1658C}  \\
                        \hline
                        CMB Dipole      & -          & - &  263.99$^{\circ}$$\pm$0.14 & 48.26$^{\circ}$$\pm$0.03 & \citet{1996ApJ...470...38L}  \\ \hline
                        Velocity Flows   & -          & - &  282$^{\circ}$ &  6$^{\circ}$ &  \citet{2009MNRAS.392..743W}, \citet{2009MNRAS.392..743W}, \citet{2010MNRAS.407.2328F}  \\ \hline
                        Quasar Alignment & -          & - &  267$^{\circ}$ &  69$^{\circ}$ & \citet{2001AA...367..381H}, \citet{2005AA...441..915H}, \citet{2011ASPC..449..441H} \\ \hline
                        CMB Octopole     & -          & - &  308$^{\circ}$ &  63$^{\circ}$ & \citet{2004MNRAS.355.1283B}  \\ \hline
                        CMB Quadrupole   & -          & - &  240$^{\circ}$ &  63$^{\circ}$ & \citet{2004MNRAS.355.1283B}, \citet{2010MNRAS.403.1739F}  \\ \hline
                        $\bigtriangleup$$\alpha/\alpha$ & -   & - &  330$^{\circ}$ & -13$^{\circ}$ & \citet{2011PhRvL.107s1101W} , \citet{2012MNRAS.422.3370K} \\ \hline
                        SPARC Galaxies & -    & HC(max) &  175.5$^{\circ}$ & -6.5$^{\circ}$ & \citet{2017ApJ...847...86Z}  \\ 
                        & -    & HC(submax) &  114.5$^{\circ}$ & 2.5$^{\circ}$ & \citet{2017ApJ...847...86Z}  \\ 
                        & -    & DF &  171$^{\circ}$ &  -15$^{\circ}$ & \citet{2018ChPhC..42k5103C}  \\ \hline      
                        Galaxy Cluster    & $\rm \Lambda$CDM  & - &  303$^{\circ}$ &  -27$^{\circ}$ & \citet{2020AA...636A..15M}  \\ \hline 
                        Infrared galaxies  &  - & - & 310$^{\circ}$  & -15$^{\circ}$  &  \citet{2014MNRAS.445.L60Y}  \\ 
                        &  - & HC & 323$^{\circ}$  & -5$^{\circ}$  &  \citet{2017MNRAS...464...768B}  \\
                        \hline   
                        Pantheon+1421Quasar & $\rm \Lambda$CDM   & HC &  316.08$^{\circ}$$^{+27.41}_{-129.48}$  &  4.53$^{\circ}$$^{+26.29}_{-64.06}$ & this paper  \\ 
                        & $\rm \Lambda$CDM   & DF &  327.55$^{\circ}$$\pm$32.45 &  51.01$^{\circ}$$\pm$26.50 & this paper  \\ \hline \hline

                \end{supertabular}
        \end{spacing}
\end{table}

%%%%%%%%%%%%%%%%%%%%%%%%%%%%%%%%%%%%%%%%%%%%%%%%%%
%Tab.2

\begin{landscape}
        \begin{table}
                \caption{Redshift tomography results using dipole fitting method for the SN-Q sample.}
                \label{T1}       % Give a unique label
                %       \begin{spacing}{1.19}
                \centering
                \begin{spacing}{2.19}
                        \begin{tabular}{|c|c|c|c|c|c|c|c|c|c|}
                                \hline
                                & $N_{\textnormal{SN+Quasar}} $& $\Omega_{m}$ & $\delta$ &  $\gamma$ & $\beta$ & $l$ ($^{\circ}$) & $b$ ($^{\circ}$) & $A$ & $B$  \\ \hline 
                                $z<0.25$ & 21 + 530 & 0.283$^{+0.049}_{-0.073}$ & 0.295$^{+0.05}_{-0.06}$ & 0.628$^{+0.098}_{-0.105}$ & 7.747$^{+3.52}_{-4.03}$ & 279.88$^{\circ}$$^{+12.15}_{-26.71}$ & 23.76$^{\circ}$$^{+25.29}_{-16.94}$ & -8.43$^{+3.32}_{-6.59}$e-4 & -2.84$^{+3.19}_{-5.11}$e-4   \\ \hline
                                
                                $z<0.50$ & 136 + 832 & 0.273$^{+0.024}_{-0.035}$ & 0.261$^{+0.017}_{-0.022}$ & 0.576$^{+0.044}_{-0.065}$ & 9.007$^{+1.36}_{-1.83}$ & 306.98$^{\circ}$$^{+20.96}_{-20.16}$ & 30.95$^{\circ}$$^{+34.52}_{-16.01}$ & -3.98$^{+3.01}_{-5.14}$e-4 & -2.20$^{+2.52}_{-4.30}$e-4  \\ \hline
                                
                                $z<0.75$ & 379 + 949 & 0.285$^{+0.018}_{-0.027}$ & 0.237$^{+0.008}_{-0.012}$ & 0.615$^{+0.027}_{-0.037}$ & 7.829$^{+0.079}_{-1.12}$ & 319.09$^{\circ}$$^{+30.80}_{-38.48}$ & 40.33$^{\circ}$$^{+29.86}_{-18.70}$ & -4.58$^{+2.10}_{-3.90}$e-4 & -1.40$^{+2.17}_{-3.73}$e-4 \\ \hline
                                
                                $z<1.00$ & 664 + 1025 & 0.285$^{+0.014}_{-0.020}$ & 0.238$^{+0.006}_{-0.009}$ & 0.633$^{+0.019}_{-0.026}$ & 7.302$^{+0.56}_{-0.81}$ & 321.78$^{\circ}$$^{+26.27}_{-37.19}$ & 52.78$^{\circ}$$^{+23.41}_{-13.84}$ & -4.65$^{+1.75}_{-3.14}$e-4 & -1.32$^{+2.02}_{-3.26}$e-4  \\ \hline
                                
                                $z<1.50$ & 1020 + 1042 & 0.289$^{+0.013}_{-0.019}$ & 0.235$^{+0.005}_{-0.008}$ & 0.633$^{+0.013}_{-0.019}$ & 7.294$^{+0.42}_{-0.56}$ & 324.72$^{\circ}$$^{+27.61}_{-23.42}$ & 45.83$^{\circ}$$^{+24.09}_{-14.43}$ & -4.67$^{+1.64}_{-3.19}$e-4 & -9.40$^{+19.5}_{-32.1}$e-5 \\ \hline
                                
                                $z<2.50$ & 1334 + 1048 & 0.293$^{+0.013}_{-0.019}$ & 0.233$^{+0.0046}_{-0.0068}$ & 0.630$^{+0.0010}_{-0.014}$ & 7.367$^{+0.309}_{-0.425}$ & 322.44$^{\circ}$$^{+34.89}_{-31.42}$ & 47.40$^{\circ}$$^{+24.75}_{-13.72}$ & -4.47$^{+1.64}_{-3.06}$e-4 & -5.30$^{+19.1}_{-32.0}$e-5 \\ \hline

                                $z<3.50$ & 1416 + 1048 & 0.294$^{+0.0051}_{-0.019}$ & 0.232$^{+0.0019}_{-0.0066}$ & 0.641$^{+0.0034}_{-0.0124}$ & 7.059$^{+0.104}_{-0.375}$ & 325.38$^{\circ}$$^{+32.01}_{-35.23}$ & 50.93$^{\circ}$$^{+24.69}_{-10.63}$ & -4.12$^{+0.66}_{-2.97}$e-4 & -3.10$^{+0.76}_{-3.15}$e-5  \\ \hline
                                
                                $z<4.20$ & 1421 + 1048 & 0.292$^{+0.015}_{-0.017}$ & 0.231$^{+0.005}_{-0.006}$ & 0.640$^{+0.009}_{-0.011}$ & 7.041$^{+0.28}_{-0.34}$ & 327.21$^{\circ}$$^{+28.24}_{-27.50}$ & 48.67$^{\circ}$$^{+8.99}_{-8.44}$ & -4.36$^{+1.98}_{-2.74}$e-4 & -6.60$^{+21.5}_{-28.6}$e-5  \\ \hline
                        \end{tabular}
                \end{spacing}
        \end{table}
\end{landscape}

\twocolumn

\end{document}